\documentclass[11pt,aps,prd,nofootinbib,superscriptaddress,preprintnumbers]{revtex4}
\usepackage{graphicx,epsfig}
\usepackage{amsmath}
\newcommand{\bea}{\begin{eqnarray}}
\newcommand{\beq}{\begin{equation}}
\newcommand{\eea}{\end{eqnarray}}
\newcommand{\eeq}{\end{equation}}

\newcommand{\Frac}[2]{\frac{\displaystyle{#1}}{\displaystyle{#2}}}
\newcommand{\lsim}{\raise0.3ex\hbox{$\;<$\kern-0.75em\raise-1.1ex\hbox{$\sim\;$}}}
\newcommand{\gsim}{\raise0.3ex\hbox{$\;>$\kern-0.75em\raise-1.1ex\hbox{$\sim\;$}}}
\newcommand{\eq}[1]{Eq.~(\ref{#1})}
\newcommand{\unity}{{\hbox{1\kern-.8mm l}}}
\newcommand{\LL}{{\mbox{\scriptsize LL}}}
\newcommand{\LR}{{\mbox{\scriptsize LR}}}
\newcommand{\RL}{{\mbox{\scriptsize RL}}}
\newcommand{\RR}{{\mbox{\scriptsize RR}}}
\begin{document}

\title{Electric dipole moments from flavoured CP violation in SUSY}
\author{L.~Calibbi}
\email{calibbi@sissa.it}
\affiliation{SISSA/ISAS and INFN, I-34013, Trieste, Italy.}
\affiliation{Departament de F\'{\i}sica Te\`orica and IFIC, Universitat de 
Val\`encia-CSIC, E-46100, Burjassot, Spain.}
\author{J.~Jones~P\'erez}
\email{joel.jones@uv.es}
\affiliation{Departament de F\'{\i}sica Te\`orica and IFIC, Universitat de 
Val\`encia-CSIC, E-46100, Burjassot, Spain.}
\author{O. Vives}
\email{oscar.vives@uv.es}
\affiliation{Departament de F\'{\i}sica Te\`orica and IFIC, Universitat de 
Val\`encia-CSIC, E-46100, Burjassot, Spain.}

\preprint{FTUV-08-0429, IFIC-08-24, SISSA-28/2008/EP}

\begin{abstract}
The so-called  supersymmetric flavour and CP problems are deeply related to
the origin of flavour and hence to the origin of the SM Yukawa couplings
themselves. We show that realistic $SU(3)$ flavour symmetries with spontaneous 
CP violation reproducing correctly the SM Yukawa matrices can simultaneously 
solve both problems without ad hoc
modifications of the SUSY model. We analyze the leptonic electric dipole 
moments and lepton flavour violation processes in these models. 
We show that the electron EDM and the decay $\mu \to e \gamma$
are naturally within reach of the  proposed experiments if the sfermion masses 
are measurable at the LHC. 

\end{abstract}

\maketitle
\section{Introduction}

The so-called Supersymmetry (SUSY) flavour and CP problems are usually taken
as the main naturalness problems SUSY has to face up \cite{Masiero:2001ep,Masiero:2005ua}. 
The formulation of the
supersymmetric flavour problem is well known: since we have no information
regarding the structure of SUSY soft-breaking terms, we could in principle
expect that entries in a soft-mass matrix are all of the same order. In
particular, this can happen in the basis where the Yukawa couplings are 
diagonal. In such a situation, FCNC and flavour-dependent CP violation
observables would receive too large contributions from loops involving SUSY
particles to satisfy the stringent phenomenological bounds on these
processes \cite{Masiero:2001ep,Masiero:2005ua}.
We can formulate the SUSY CP Problem in a similar way. If CP is not a
symmetry of the model we naturally expect all complex parameters in the
model to have $O(1)$ phases. In this case, the phases in
flavour-independent
terms typically generate too
large contributions to the so-far unobserved electric dipole moment
(EDM) of the electron and neutron \cite{Pospelov:2005pr,Raidal:2008jk}.

The basis of both problems lies clearly on our total ignorance about the 
origin of the observed flavour and CP-violation in our theory. However, 
notice that these problems are not restricted to supersymmetry. Even the 
Standard Model (SM) shares the flavour problem with SUSY in exactly the same terms.
If we had not measured the quark and lepton masses and mixings we would
naturally expect all the elements in the Yukawa matrices to be $O(1)$. Yet, if
one gives such a structure to the Yukawa matrices, the predicted fermion
masses and mixings would never agree with the observed ones. Therefore, we
have to conclude that there is a much stronger flavour problem in 
the SM than in the MSSM. The real flavour problem is simply our
inability to understand the complicated structures in the quark and lepton
Yukawa couplings, and likewise for the soft-breaking flavour structures in the
MSSM. On the other hand, there seems to be no  direct analog of the SUSY CP
problem in the SM. In fact, the phases in the SM Lagrangian are
already $O(1)$ without being in conflict with the experimental measurements.
However, this apparent ``fact'' is also misleading. Notice that, due to the particle
content of the SM, the only complex 
parameters in the Lagrangian are the Yukawa couplings themselves and we
have measured them to be small. Once more, if we had not known the fermion
masses and mixings beforehand and wrote arbitrary complex Yukawa parameters,
we would also have a severe SM CP problem. Since the SUSY CP
problem is basically due to the flavour-independent phases in the MSSM,   
both facts can suggest the idea that the flavour and the CP problems are 
indeed related and solving the flavour problem  while restricting the 
CP phases to the flavour sector would also solve the CP problem. 

A particularly attractive solution to these problems (both in the SM and in
SUSY) is found on models based on flavour
symmetries. In these models, the flavour 
structure
of the Yukawa matrices is only  generated after the breaking of a flavour 
symmetry~\cite{Ross:2000fn,Froggatt:1978nt,Leurer:1992wg,Dine:1993np,Kaplan:1993ej,Pomarol:1995xc,Barbieri:1995uv,Binetruy:1996xk,Dudas:1996fe,Plentinger:2008nv}
and the flavour structure of the SUSY soft-breaking terms would also
originate through the same mechanism \cite{Nir:1993mx,Leurer:1993gy,Nir:1996am,Ross:2004qn,Joshipura:2000sn}. Thus, finding a solution to the 
SM flavour problem will generally solve at the same time the 
so-called SUSY flavour problem to a sufficient degree, although 
probably still allowing naturally suppressed
contributions that might bring more information about the flavour sector. 
Regarding the SUSY CP problem, if we want to restrict all CP
phases in SUSY to the flavour sector, this can be achieved by
postulating an exact CP symmetry spontaneously broken
in the flavour sector. This would remove all
flavour-independent phases, but still produce interesting
 observable sources of CP violation.

In this work we shall analyze how a flavour symmetry can solve the flavour and
CP problems in a SUSY scenario. We shall also distinguish typical signatures
of such a symmetry in the lepton sector\footnote{Observables in the quark sector, such as the neutron EDM, shall be studied in a future work~\cite{WIP}}. In Section~\ref{sec:flavmodel} we shall analyze a general flavour symmetry model which can solve both problems, based on~\cite{Ross:2004qn}. In Section~\ref{sec:EDMsection} we shall identify the most important flavour-dependent contributions to leptonic EDMs, in order to quantify the expected amount of CP violation in such models. In Section~\ref{sec:MuegSection} we shall study in a similar manner the relevant lepton flavour violation (LFV) processes. Finally, in Section~\ref{sec:numerical} we will show the regions of the SUGRA parameter space that are sensitive to future EDM and LFV experiments.

\section{Flavour symmetries and spontaneous CP breaking}
\label{sec:flavmodel}

Following the original ideas of Froggatt-Nielsen~\cite{Froggatt:1978nt}, flavour symmetries explain the peculiar structure of the SM Yukawa couplings as the result of a spontaneously broken symmetry associated with flavour. The three generations of SM fields are charged under this symmetry such that the SM Yukawa couplings are not allowed in the limit of exact symmetry. One or several scalar vacuum expectation values (vevs) breaking this symmetry must be inserted in a non-renormalizable operator, suppressed by a heavy mediator mass, to compensate the charges. If the scalar vev is smaller than the mediator scale, this provides a small expansion parameter that can be used to explain the hierarchy of the observed Yukawa couplings.

In the context of a supersymmetric theory, an unbroken flavour symmetry would apply equally to the fermion and scalar sectors. This implies that in the limit of exact symmetry the soft-breaking scalar masses and the trilinear couplings must be invariant under the flavour symmetry. This has different implications in the case of the scalar masses and the trilinear couplings.

The scalar masses are couplings $\phi \phi^\dagger$, thus flavour-diagonal
couplings are clearly invariant under any symmetry, i.e. diagonal soft-masses
are always allowed by the flavour symmetry. Therefore diagonal scalar masses
will be of the order of the SUSY soft breaking scale. However, in general,
this does not guarantee that they are family universal. Being universal or not
will depend on the considered family symmetry. In the case of an Abelian
\cite{Froggatt:1978nt,Leurer:1992wg,Pomarol:1995xc,Binetruy:1996xk,Dudas:1996fe,Dreiner:2003yr,Kane:2005va,Chankowski:2005qp} family symmetry, the symmetry does not relate different 
generations and
therefore diagonal masses can be different. In this case, Flavour Changing
Neutral Current (FCNC) and CP violation phenomenology set very strong
constraints on the differences between these flavour diagonal masses and
Abelian family symmetries 
have serious difficulties to satisfy these
constraints~\cite{Nir:2002ah}. On the other hand, a non-Abelian family symmetry groups two or
three generations in a single multiplet with a common mass, thus solving the
FCNC problem. This was one of the main motivations for the construction of the
first $SU(2)$ flavour models~\cite{Barbieri:1995uv,Barbieri:1996ww}, where the 
first two
generation sfermions, facing the strongest constraints, share a common mass.
In the case of an $SU(3)$ flavour symmetry \cite{King:2001uz,King:2003rf,Ross:2004qn,deMedeirosVarzielas:2005ax}, all three generations have the
same mass in the unbroken family symmetry limit. For this reasons, in the
following we will consider non-Abelian family symmetries, and more precisely
$SU(3)$ flavour symmetries. 

 On the contrary, trilinear couplings are
completely equivalent to the Yukawa couplings from the point of view of the
symmetry because they involve exactly the same fields (scalar or fermionic
components). Thus, they are forbidden by the symmetry (with the possible
exception of the $(3,3)$ component in $SU(2)$ models) and generated only after
symmetry-breaking as a function of small vevs.

In addition to the renormalizable mass operators in the Lagrangian, we can
construct non-renormalizable operators neutral under the flavour symmetry
inserting an appropriate number of flavon fields. The flavon fields, charged
under the symmetry, are responsible for the spontaneous symmetry breaking
once they acquire a vev. Then, higher dimensional operators involving two SM
fermions and a Higgs field, with several flavon vevs suppressed by a large mediator mass, generate the observed Yukawa couplings. In the same way, these flavon fields will couple to the scalar fields in all possible ways allowed by the symmetry and, after spontaneous symmetry breaking, they will generate a non-trivial flavour structure in the soft-breaking parameters. Therefore, by being generated by insertions of the same flavon vevs, we can expect the structures in the soft-breaking matrices and the Yukawa couplings to be related. Our starting point in our analysis of the soft-breaking terms must then involve an analysis of the texture in the Yukawas, in order to reproduce first the correct masses and mixings.

To fix the Yukawa couplings, we accept that the smallness of CKM mixing angles is due to the smallness of the off-diagonal elements in the Yukawa matrices with respect to the corresponding diagonal elements, and we make the additional simplifying assumption of choosing the matrices to be symmetric. With these two theoretical assumptions,  and using the ratio of masses at a high scale to define the expansion parameters in the up and down sector as $\bar \varepsilon= \sqrt{m_s/m_b}$ and $\varepsilon= \sqrt{m_c/m_t}$, we can fix the Yukawa textures in the quark sector to be:
\begin{eqnarray}  
\label{fit}
Y_d\propto\left( 
\begin{array}{ccc}
0 & b\ \bar \varepsilon^{3} &c\  {\ \bar \varepsilon^{3}} \\ 
b\ \bar \varepsilon^{3} & {\bar \varepsilon^{2}} & 
a \ {\ \bar \varepsilon^{2}} \\ 
c\ {\ \bar \varepsilon^{3}} & a \ {\ \bar \varepsilon^{2}} & 1%
\end{array}%
\right),~~~~~~ Y_u\propto \left( 
\begin{array}{ccc}
0 & b^\prime {\ \varepsilon^{3}} & c^\prime {\ \varepsilon^{3}} \\ 
 b^\prime{\ \varepsilon^{3}} & {\ \varepsilon^{2}} & 
a^\prime \varepsilon^{2} \\ 
 c^\prime{\ \varepsilon^{3}} & a^\prime \varepsilon^{2} & 1%
\end{array}
\right ) \, ,
\end{eqnarray}
with $\bar \varepsilon\simeq 0.15$, $\varepsilon \simeq 0.05$, $b=1.5$, $%
a=1.3$, $c=0.4$ and $a^\prime,b^\prime,c^\prime$ are poorly fixed from
experimental data \cite{Roberts:2001,Ross:2004qn,Kane:2005va}. 
Unfortunately, the Yukawa couplings in
the leptonic sector can not be determined from the available phenomenological
data. The left-handed neutrino masses and mixings cannot unambiguously fix 
the neutrino Yukawa couplings in a seesaw mechanism. Therefore only the
charged lepton masses provide useful information on leptonic Yukawas.
For simplicity, we choose to work in a Grand Unified model at high scales,
a possibility which is favored by the unification of the bottom and tau
Yukawa couplings. In this case, charged lepton and down-quark
(and the neutrino and up-quark) flavour matrices are the same except for the 
different vev of a Georgi-Jarlskog
Higgs field \cite{Georgi:1979df} to unify the second and first generation masses. 

With this Yukawa structure as our starting point, we will generate the flavour
structure of the soft-breaking terms in an explicit example based in an
$SU(3)$ flavour symmetry. Under this symmetry, the three generations of SM
fields, including both $SU(2)_L$-doublets and singlets, are triplets ${\bf 3}$ and the
Higgs fields are singlets. Therefore Yukawa couplings and trilinear terms:
${\bf 3} \times {\bf 3} \times {\bf 1}$ are not allowed by the $SU(3)$. In the
theory, we have several flavon fields, that we call $\theta_3$, $\theta_{23}$
(anti-triplets ${\bf \bar 3}$), $\bar \theta_3$ and $\bar \theta_{23}$
(triplets ${\bf 3}$). The symmetry is broken in two steps, first $\theta_3$
and $\bar \theta_3$ get a vev, $\propto (0,0,1)$, breaking $SU(3)$ into
$SU(2)$.  Subsequently a smaller vev of $\theta_{23}$ and $\bar \theta_{23}$,
$\propto (0,1,1)$, breaks the remaining symmetry \cite{King:2001uz}.

To reproduce the Yukawa textures, the large third generation Yukawa couplings
require a $\theta_3$ (and $\bar \theta_{3}$) vev of the order of the mediator
scale, $M_f$ (slightly smaller in the up sector and $\tan
\beta$-dependent in the down sector as shown in the appendix), while $\theta_{23}/M_f$ (and $\bar \theta_{23}/M_f$) have vevs
of order $\varepsilon$ in the up sector and $\bar \varepsilon$ in the down
sector, with different mediator scales in both sectors. In this way, third
generation Yukawa couplings are generated by $\theta_{3}^{i}\theta _{3}^{j}$,
while couplings in the 2--3 block of the Yukawa matrix are always given by
$\theta _{23}^{i}\theta _{23}^{j}
\left(\theta_3 \overline{\theta}_3\right)$\footnote{Notice that we add the
  scalar product $(\theta_3\overline{\theta}_3)$ to this operator with
  respect to Refs.~\cite{King:2001uz,King:2003rf,Ross:2004qn} to be able to
  generalize to different $\tan
  \beta=(\langle\theta_3^u\rangle/\langle\theta_3^d\rangle)^2$ values.} (with
possibly the Georgi-Jarlskog field $\Sigma$ to unify quark and leptonic Yukawa couplings, 
not included here for simplicity, see Refs.~\cite{King:2001uz,King:2003rf,Ross:2004qn}).  The
couplings in the first row and column of the Yukawa matrix are given by
$\epsilon ^{ikl}\overline{\theta }_{23,k}\overline{\theta }_{3,l} \theta
_{23}^{j}\left( \theta _{23}\overline{\theta_{3}}\right)^n $ where to
reproduce the texture in Eq.~(\ref{fit}) we must force $n=1$. Unfortunately, the
$SU(3)_{fl}$ symmetry is not enough to reproduce the textures in
Eq.~(\ref{fit}) and we must impose some additional global symmetries
(typically $Z_N$ symmetries) to guarantee the correct power structure and to
forbid unwanted terms, like a mixed $\theta_{3}^{i}\theta _{23}^{j}$ term, in
the effective superpotential. The basic structure of the Yukawa superpotential
(for quarks and leptons) is then given by
\begin{eqnarray}
W_{\rm Y} &=& H\psi _{i}\psi _{j}^{c} \left[\theta _{3}^{i}  \theta
_{3}^{j}+\theta _{23}^{i} \theta _{23}^{j}
\left(\theta_3\overline{\theta}_3\right)  + 
\epsilon ^{ikl} \overline{ \theta}_{23,k} {\overline{ \theta }_{3,l}} \theta _{23}^{j}\left(
 \theta _{23} {\overline{\theta} _{3}}\right) +\dots \right],
\end{eqnarray} 
where to simplify the notation, we have normalized the flavon fields to the corresponding mediator mass, i.e., all the flavon fields in this equation should be understood  as $\theta_i/M_f$. 
This structure is quite general for the different $SU(3)$ models we can build, and for additional details we refer to \cite{King:2001uz,King:2003rf,Ross:2004qn}.

In the same way, after $SU(3)$ breaking the scalar soft masses deviate from
exact universality. As explained above, $\phi _{i}^{\dagger }\phi _{i}$ is
completely neutral under gauge and global symmetries and gives rise to a
common contribution for the family triplet. However, after $SU(3)$ breaking,
terms with additional flavon fields give rise to important corrections
\cite{Ross:2002mr,Ross:2004qn,Antusch:2007re,Olive:2008vv}. Any invariant
combination of flavon fields can also contribute to the sfermion masses. In
this case, it is easy to see that the following terms will always contribute
to the sfermion mass matrices \footnote{Discrete non-Abelian subgroups of
  $SU(3)$ \cite{Babu:2002dz,Hirsch:2003dr,Altarelli:2005yx,deMedeirosVarzielas:2005qg,Ma:2006ip,Ma:2006sk,deMedeirosVarzielas:2006fc,Feruglio:2007uu} would have a  similar leading order structure in the soft 
mass matrices.}:
\begin{align}
\label{eq:minimal}
(M^2_{\tilde f})^{ij} = m_0^2 \bigg(\delta ^{ij} &
  +\frac{\displaystyle{1}}{\displaystyle{M_f^{2}}}\left[\theta _{3, i}^{\dagger}
\theta _{3,j}  + \overline{ \theta} _{3}^i \overline{ \theta} _{3}^{j\dagger} + 
\theta _{23, i}^{\dagger }\theta_{23,j} + \overline{ \theta} _{23}^i 
\overline{ \theta}_{23}^{j\dagger} \right] \nonumber \\
 & + \Frac{1}{M_f^4}(\epsilon ^{ikl}\overline{\theta }_{3,k}
\overline{\theta }_{23,l})^{\dagger }(\epsilon ^{jmn}
\overline{\theta }_{3,m}\overline{\theta }_{23,n})+\quad\ldots\quad\bigg),
\end{align}
where $f$ represents the $SU(2)$ quark and lepton doublets or the up
(neutrino) and down (charged-lepton) singlets. Notice
that we have three different mediator masses, $M_f=M_L, M_u, M_d$, because 
the flavour symmetry must commute with the SM symmetry and therefore the 
vector-like mediator fields must have the SM quantum numbers of the usual 
particles\footnote{Nevertheless, in all our numerical calculations below we take only two
different mediator mass $M_d$ and $M_u=M_L$ for simplicity}. 
With this terms, we can see some similarities between the
flavour structure of the Yukawa and soft-mass matrices. In particular, the
offdiagonal (2,3) elements in the Yukawa matrix is given by $\overline{
  \theta} _{23}^{2} \overline{ \theta}_{23}^{3}/M_f^2\simeq \epsilon_f^2$
(the term $(\theta_3 \overline \theta_3)$ factorizes from all the
contributions in the Yukawa matrix), while
    in the soft masses it is $\overline{ \theta} _{23}^{i\dagger } \overline{
        \theta}_{23}^{j} /M_f^2$, also of order $\epsilon_f^2$. Still, it is
        possible to build other invariant combinations with different flavon
        fields that can not be present in the superpotential. This is due to
        the fact that the superpotential must be holomorphic, i.e. can not
        include daggered fields, while the soft masses, coming from the
        K\"ahler potential, are only a real combination of fields. In some
        cases, the global symmetries can allow a combination like $\theta
        _{23,i} \overline{ \theta}_{3}^{j} + h.c.$ to the soft masses, even
        though the combination $\theta _{23,i}\theta_{3,j}$ is forbidden by
        global symmetries in the superpotential. This structures would affect
        strongly the phenomenology of third generation physics \cite{WIP}. 
In any case,
        although possible, this situation is very rare and usually the
        structure of the soft terms follow the Yukawa structure. 
It is important to emphasize at this point that these deviations from 
universality in the soft-mass matrices
proportional to flavour symmetry breaking come always
through corrections in the K\"ahler potential.
Therefore, these effects will be important only in
gravity-mediation SUSY models where the low-energy soft-mass matrices are
mainly generated through the K\"ahler potential \cite{Brignole:1993dj}. In other mediation
mechanism, as gauge-mediation \cite{Giudice:1998bp} or anomaly mediation
\cite{Randall:1998uk,Giudice:1998xp,Pomarol:1999ie}, where these K\"ahler
contributions to the soft masses are negligible, flavour effects in the soft
mass matrices will be basically absent.

In the case of the trilinear couplings we  have to emphasize that from the
point of view of the flavour symmetry these couplings are completely
equivalent to the corresponding Yukawa coupling. This means that they
necessarily involve the same combination of flavon vevs, although order one
coefficients are generically different because they require at least an
additional coupling to a field mediating SUSY breaking (in general coupled in
different ways in the various contributions). Therefore, from our point of view, we expect that the trilinear couplings have the same structure as the Yukawa matrices in the flavour basis. However in general they are {\bf not} proportional to the Yukawas, because of different $O(1)$ coefficients in the different elements. Thus, we can expect that going to the SCKM basis does not diagonalize the trilinear matrices. In fact, the trilinear matrices maintain the same structure as in the flavour basis and only the $O(1)$ coefficients are modified.

We should now take into account, specially in a gravity-mediated scenario, the
canonical normalization of the kinetic terms (K\"ahler potential). 
However as shown in Refs.~\cite{King:2004tx,Antusch:2007re} these canonical 
transformations 
do not modify the general structure of the different flavour matrices and 
change only the unknown $O(1)$ coefficients. As, at this level, these 
coefficients can not be fixed by the flavour symmetry, the previous discussion
on the different flavour matrices remains valid in the canonical basis.

We have now fixed the flavour structure that we can expect in the soft-breaking matrices. Now, we are ready to consider the problem of CP violation in these flavour matrices. In fact, in the SM, CP violation is deeply related with flavour. The only possible complex parameters in the SM are the Yukawa couplings themselves and all observed CP violation is consistent with a single observable phase in the CKM matrix.  Therefore we need complex flavon vevs to reproduce the
observed CP in the CKM mixing matrix.
\begin{table}
\begin{center}
\begin{tabular}{|c||c|c|}
\hline
EDM & Current Bound (e cm) & Future Bound (e cm) \\
\hline \hline
$|d_e|$ & $\leq 1.4\times10^{-27}$~\cite{CurreEDM} & $\sim10^{-32}$~\cite{FuteEDM}\\
$|d_\mu|$ & $\leq 7.1\times10^{-19}$~\cite{CurrmEDM} & $\sim10^{-23}$~\cite{FutmEDM}\\
$|d_\tau|$ & $\leq 2.5\times10^{-17}$~\cite{CurrtEDM} & $\sim10^{-20}$~\cite{FuttEDM}\\
$|d_n|$ & $\leq 2.9\times10^{-26}$~\cite{CurrnEDM} & $\sim10^{-28}$~\cite{FutnEDM}\\
\hline
\end{tabular}
\end{center}
\caption{\label{Table:EDMs} Current constraints to EDMs (left) and reach of future experiments (right).}
\end{table}
On the other hand, the Supersymmetric CP problem concerns the fact that, in
the presence of flavour independent phases in the $\mu$ term and trilinear
couplings, Supersymmetry gives rise to contributions to the EDMs at 1 loop
order with no suppression associated to flavour \cite{Ellis:1982tk,Buchmuller:1982ye,Nath:1991dn, Franco:1983xm,Polchinski:1983zd,Dugan:1984qf,Fischler:1992ha,Ayazi:2007kd}. These one-loop 
contributions,
for SUSY masses below several TeV, can easily exceed the present experimental
bounds shown on Table~\ref{Table:EDMs} by one or two orders of magnitude. This
fact forces most models to demand unnatural requirements, such as vanishing
phases, very large mediator masses or engineered cancellations between
different contributions to the process~\cite{Ibrahim:1998je}. 

In this aspect, flavour symmetries with spontaneous breaking of CP provide an
interesting solution. CP is an exact symmetry of the theory before the
breaking of the flavor symmetry. Thus, above the scale of flavour breaking,
all terms in the K\"ahler potential, which gives rise to the soft masses and
the $\mu$ term (by the Giudice-Masiero mechanism~\cite{Giudice:1988yz}), are
real. Even after the breaking of flavour and CP symmetries, $\mu$ receives flavour-suppressed complex corrections only at the two-loop level~\cite{Barr:1988wk}. Finally, trilinear terms are only generated after symmetry-breaking with the same phase structure as the Yukawa couplings, and it can be proven that diagonal elements in $A_{ij}$ are real at leading order in the SCKM basis~\cite{Ross:2004qn}. In this way, the Supersymmetric CP problem is naturally solved. Nevertheless, flavour-dependent phases do exist and can contribute to the fermion EDMs as we will see below.

In the following sections, even though we use exact expressions to calculate
all our observables, we shall use the mass insertion (MI) approximation~\cite{Gabbiani:1996hi, Hagelin:1992tc} to
quantify and explain all important contributions (as
in~\cite{Masina:2002mv}). In terms of the slepton mass matrix terms, these MIs are defined as:
\begin{align}
\label{DeltaLR}
(\delta^e_{\LR})_{ij}  =  \frac{v_d}{\sqrt{2}M^2_{\tilde e}}[A_{e,ij}^* -& \delta_{ij} Y_i \mu\tan\beta] = (\delta^e_{\RL})^*_{ji} \nonumber \\ 
(\delta^e_{\LL})_{ij}  =  \frac{(m^2_{\tilde L})_{ij}}{M^2_{\tilde e}} \qquad & \qquad
(\delta^e_{\RR})_{ij}  =  \frac{(m^2_{\tilde e})_{ji}}{M^2_{\tilde e}}
\end{align}
where $M^2_{\tilde e}$ is the average slepton mass. From the soft slepton mass matrices of Eq.~(\ref{soft0}), we can read the approximate MIs in the $\mu-e$, $\tau-e$ and $\tau-\mu$ sector respectively:
\begin{subequations}
\label{DetailMI}
\begin{eqnarray}
(\delta^e_{\LL})_{12} \approx {\epsilon^2 \bar{\epsilon} \over 3} ;\;\;\; 
 & (\delta^e_{\RR})_{12} \approx \Frac{\bar{\epsilon}^3}{3} ;\;\;\; &
(\delta^e_{\LR})_{12} \approx A_0 {m_{\tau} \over M_{\tilde e}^2}\bar{\epsilon}^3 \\ 
(\delta^e_{\LL})_{13} \approx {\bar\epsilon^{3}~y_{33}^\nu } ;\;\;\; 
& (\delta^e_{\RR})_{13} \approx \Frac{\bar{\epsilon}^3}{3} ;\;\;\; &
(\delta^e_{\LR})_{13} \approx A_0 {m_{\tau} \over M_{\tilde e}^2} \bar{\epsilon}^3 \\
(\delta^e_{\LL})_{23} \approx \bar\epsilon^{2}~y_{33}^\nu ;\;\;\; 
& (\delta^e_{\RR})_{23} \approx \bar{\epsilon}^{2} ;\;\;\; &
(\delta^e_{\LR})_{23} \approx A_0 {m_{\tau} \over M_{\tilde e}^2} 3 \bar{\epsilon}^2 
\label{delta23}
\end{eqnarray}
\end{subequations}
where we have not included the renormalization group running from the
unification scale down to low energy. Such running effects can be sizeable in
the LL and LR sectors, due to the presence of heavy RH neutrinos with large
Yukawa couplings \cite{Borzumati:1986qx,Casas:2001sr,Masiero:2002jn}. Moreover, in the present case there
are new contributions to the running given by the non-universality of the
soft-mass and trilinear matrices. These effects can be important even in the
RR sector.
As an example, the element $\left(m^2_{\tilde e}\right)_{32}$ of
the RH slepton mass matrices gets the following RG correction in SCKM basis:
\begin{eqnarray}
\left(m^2_{\tilde e}\right)_{32}(M_{\rm SUSY})
&\simeq & \left[\left(m^2_{\tilde e}\right)_{32}(M_{U}) \left(1 - {1 \over 16 \pi^2}
\left(2 y_\tau^2 \right) \right)  -
4{ A^e_{33} A^{e \dagger}_{32} \over 16 \pi^2} \right]
\log\left({M_U \over M_{\rm SUSY}} \right)
\end{eqnarray}
where $M_U$ is the unification scale, $M_{\rm SUSY}$ the average SUSY
mass and the matrix elements are evaluated at $M_U$ using the SCKM expressions given in the
Appendix. In the SCKM basis we have $A^e_{13}/A_0 \sim (Y_\nu)_{13} \sim
\mathcal{O}(\bar{\epsilon}^3)$, $A^e_{23}/A_0 \sim (Y_\nu)_{23} \sim
\mathcal{O}(\bar\epsilon^2)$, $A^3_{33}/A_0 \sim m_\tau/(v \cos \beta)$
and $(Y_\nu)_{33} \sim \mathcal{O}(1)$, and $y_e \sim \mathcal{O}(\bar\epsilon^4)$,
$y_\mu \sim \mathcal{O}(\bar\epsilon^2)$, $y_\tau \sim \mathcal{O}(1)$.
Therefore, we expect sizeable correction from the running only in the
$\tau-\mu$ and $\tau-e$ sectors, i.e. where the third generation is involved.

\section{EDMs and Flavour Phases}
\label{sec:EDMsection}

Fermion EDMs, $d_\psi$, are induced through effective dimension-six operators (with an implicit Higgs insertion providing the chirality charge) of the form:
\begin{equation}
{\cal L} = -\frac{d_\psi}{2}\left[\bar\psi\sigma^{\mu\nu}\gamma^5\psi\right] F_{\mu\nu},
\end{equation}
being related to the imaginary part of a chirality-changing, flavour-diagonal loop process. In the SM these processes are greatly
suppressed: the prediction for the electron EDM is lower than $O(10^{-40})$ e~cm~\cite{SMeEDM}. This makes EDMs very convenient observables where to look for new physics related to CP violation.

As we have seen above, if CP is spontaneously broken in the flavour sector,
the usual flavour-independent phases coming from $\mu$ and $A_f$ are
approximately zero and we expect EDMs to be under control\footnote{Notice
  that, since we are assuming gaugino mass unification, we can always take the
  unified gaugino mass as real, corresponding to the usual convention in the
  constrained MSSM.}. Nevertheless, flavour-dependent phases in the soft-mass matrices and trilinears can also give large contributions~\cite{Abel:2001vy,Bartl:2003ju}. In order to be sure that the SUSY CP problem is solved, it will be necessary to quantify the expected order of magnitude of the EDMs produced by $O(1)$ phases on these terms. If the current constraints are respected, we can then contrast these predictions with the expected sensitivity of future EDM experiments, shown also on Table~\ref{Table:EDMs}.

In order to identify the dominant terms in $d_e$, one needs to know the size
and phases of the different mass insertions.  In fact, observable phases will
correspond to rephasing invariant combinations of mass insertions and Yukawa
elements~\cite{Botella:2004ks}. Even in the general $SU(3)$ framework presented in the previous
section, these combinations depend strongly on the particular model one 
takes into account. Thus, as
a first step, it is of interest to study the magnitude of each potential
contribution to $d_e$ assuming a generic phase of order one for the whole
rephasing invariant combination, using the expected size of the offdiagonal
elements in the flavour symmetry. We will then proceed with a second analysis,
considering the explicit model by Ross, Velasco-Sevilla and Vives (RVV)
\cite{Ross:2004qn} presented in the Appendix.

One-loop MSSM contributions to charged lepton EDMs $d_l$ ($l=e,\mu,\tau$)
involve diagrams with neutralinos and
charginos~\cite{Bartl:1999bc,Bartl:2001wc}. However,
the chargino contribution only involves a flavour diagonal left-handed
sneutrino propagator and, due to hermiticity, the sensitivity to the phases within the sneutrino mass matrix is lost. With a vanishing phase for $\mu$, we can neglect the chargino contribution to $d_l$, and concentrate on neutralinos completely. Neutralino contributions to $d_e$ are:
\begin{eqnarray}
\label{eEDM}
d_e^{\chi^0}&=&\left(\frac{e}{16\pi^2}\right) \Im m \left(A^L_{eij} A^R_{eij}\right)
\frac{1}{m_{\chi^0_j}} F\left(\frac{m^2_{\chi^0_j}}{m^2_{\tilde{l}_i}}\right)
\\
 A^L_{eij} & = & Y_e~{\cal N}^\ast_{3j}~{\cal R}_{\tilde{e}_R i}
-\frac{g^\prime}{\sqrt{2}}~{\cal N}^\ast_{1j}~{\cal R}_{\tilde{e}_L i}
-\frac{g}{\sqrt{2}}~{\cal N}^\ast_{2j}~{\cal R}_{\tilde{e}_L i} \\
 A^R_{eij} & = & Y_e~{\cal N}^\ast_{3j}~{\cal R}^\ast_{\tilde{e}_L i}
+\sqrt{2}g^\prime~{\cal N}^\ast_{1j}~{\cal R}^\ast_{\tilde{e}_R i}
\end{eqnarray}
with ${\cal R}_{\tilde{e} i}$ and ${\cal N}_{kj}$ being elements of the
charged slepton and neutralino mixing matrices \cite{Bartl:1999bc}, respectively and the loop function $F(x)=\frac{x}{2(x-1)^3}\left(x^2-1-2x\log x\right)$.

In terms of mass insertions, the most relevant contributions from Eq.~(\ref{eEDM}) can be written as:
\begin{equation}
\frac{d_e}{e} = \frac{\alpha M_1 }{8\pi \cos^2\theta _W M^2_{\tilde{e}}}
\Im m \bigg[(\delta^e_{\LL})_{1i} (\delta^e_{\LR})_{i1}\, f_1
+ (\delta^e_{\LR})_{1i} (\delta^e_{\RR})_{i1}\, f_2
+ (\delta^e_{\LL})_{1i}(\delta^e_{\LR})_{ij} (\delta^e_{\RR})_{j1} f_3 \bigg] \ \ \
\label{mdmedm} 
\end{equation}
where $M^2_{\tilde e}$ is the average slepton mass and the loop functions, 
$f_i$, can be derived from~\cite{Masina:2002mv}.

Let us explain briefly why these are the relevant contributions, and identify
the dominant ones. All phases in this $SU(3)$ flavour model are contained
within the sfermion mass matrices and thus we shall need at least one mass insertion on the slepton line. Since all flavour-conserving insertions $(\delta^e_{\LR})_{ii}$ are real to leading order, we will need to combine at least two flavour-changing mass insertions, $(\delta^e_{AB})_{ij}$.

Regardless of the number of mass insertions, we shall always have two situations: one in which the neutralino line couples to the fermion through a gaugino and a higgsino, and one through two binos (although interactions with two higgsinos also contribute, they are suppressed by at least an additional Yukawa coupling, so we shall not discuss them). In the gaugino-higgsino case, the slepton line will need to maintain its handedness and again, due to the hermiticity of the full slepton mass matrix, loses all dependency on the flavon phases. Thus, $d_e$ is due entirely to diagrams with pure binos as the vertices, where a LR transition is required.

With two mass insertions, the only contribution with physical phases
comes from a combination of insertions like
$(\delta^e_{\LR})_{1i}(\delta^e_{\RR})_{i1}$ or $(\delta^e_{\LL})_{1i}(\delta^e_{\LR})_{i1}$. With three mass insertions, we consider only contributions with a single LR insertion. It is well-known that each $\delta^e_{\LR}$ insertion is suppressed by a cumulative factor $m_\tau/M_{\tilde e}$. Therefore the dominant contribution comes from $(\delta^e_{\LL})_{1i}(\delta^e_{\LR})_{ij}(\delta^e_{\RR})_{j1}$. Obviously, the largest contribution is the one that involves a central $(\delta^e_{\LR})_{33}$, due to the $m_\tau\tan\beta$ enhancement. Thus, the most important contribution to $d_e$ with three mass insertions is the pure bino $(\delta^e_{\LL})_{13}(\delta^e_{\LR})_{33}(\delta^e_{\RR})_{31}$.

Regarding whether the two or three insertion contribution dominates, it shall depend on the magnitude of $\tan\beta$ and the size of the off-diagonal terms in the trilinears. Evidently for a large enough $\tan\beta$, the three-insertions contribution shall dominate no matter the size of $(\delta^e_{\LR})_{ij}$, but for low values the situation is not so clear, especially if $A_0$ is large. Using \eq{soft0} in the Appendix, we can quantify the magnitude of each insertion as:
\begin{eqnarray}
\label{LRLR}
(\delta^e_{\LR})_{1i}(\delta^e_{\RR})_{i1} & \approx & A_0 \bar\epsilon^6
\frac{m_{\tau}}{M_{\tilde e}^2} \\
\label{LLLRRR}
(\delta^e_{\LL})_{13}(\delta^e_{\LR})_{33}(\delta^e_{\RR})_{31} & \approx &
\bar\epsilon^6 y_{33}^\nu \frac{m_\tau}{M_{\tilde e}}\tan\beta\, ,
\end{eqnarray}
and therefore for $A_0\simeq M_{\tilde e}$ the triple mass insertion is 
dominant except for small values of $ y^\nu_{33} \tan \beta$. %Notice also the relative sign between both terms, which means that destructive interference can take place.

It is also important to take into account that \eq{LRLR} has the same structure for the 12 and 13 elements. This means that, with a maximum phase on each element, we can double the two-insertion contribution. In any case, as these terms are all proportional to $A_0$, in order to do an appropriate study one needs to take the case where $A_0=0$ as a standard, and then understand further deviations when $A_0\neq0$.

Let us turn now to $d_\mu$. For $A_0=0$, the structure of the dominant terms for $d_\mu$ shall be quite similar to the one for $d_e$. We shall have the main contribution coming from the triple mass insertion $(\delta^e_{\LL})_{23}(\delta^e_{\LR})_{33}(\delta^e_{\RR})_{32}$, which is enhanced by $m_\tau\tan\beta$. However, due to the flavour structure of our model, we should expect a suppression of order $\bar\epsilon^4$, instead of $\bar\epsilon^6$. Thus, $d_\mu$ is about two orders of magnitude larger than $d_e$. This is similar to the usual mass scaling relation, which predicts $d_\mu$ to be larger by $m_\mu/m_e$, also two orders of magnitude.
When $A_0\neq0$, the double insertion can be relevant for low $\tan
\beta$ similarly to the case of $d_e$ analyzed in Eqs.~(\ref{LRLR}) and
(\ref{LLLRRR}). However, contrary to $d_e$, where both
$(\delta^e_{\LR})_{12}(\delta^e_{\RR})_{21}$ and
$(\delta^e_{\LR})_{13}(\delta^e_{\RR})_{31}$ are of the same order
($\bar\epsilon^6$), in this case the dominant contribution only comes from
$(\delta^e_{\LR})_{23}(\delta^e_{\RR})_{32}$, which is again of order
$\bar\epsilon^4$. 

The situation for $d_\tau$ is critically different when $A_0=0$. In this case
the $m_\tau\tan\beta$ enhancement is lost, and the main triple MI contribution
is due to
$(\delta^e_{\LL})_{32}(\delta^e_{\LR})_{22}(\delta^e_{\RR})_{23}$. This would be smaller
than $d_\mu$ by a factor $m_\mu/m_\tau$, almost two orders of magnitude, and thus 
$d_\tau$ clearly violates the naive scaling relation.
In contrast, when $A_0\neq0$, the main contributions from the double MIs with
a flavour changing $\delta_{\LR}$ insertion are identical in magnitude to
those for $d_\mu$ and so, we expect $d_\tau$ to be of size comparable
to $d_\mu$. In this case, we should take into account the possible presence of
subdominant phases in the SCKM basis in flavour diagonal trilinear couplings
\cite{Ross:2002mr}. In any case, this breaks again the mass scaling relation, even though not so drastically as in the previous situation.
The observation of such a bizarre behavior would be a very clear signal
favoring these type of flavour models.

\section{Lepton Flavour Violating decays}
\label{sec:MuegSection}

As discussed in the previous sections, supersymmetric flavour models are
characterized by non-universal scalar masses at the scale where the SUSY
breaking terms appear. Moreover, the trilinear $A_f$ matrices are in general
not aligned with the corresponding Yukawa matrices. This determines the arising of potentially
large mixing among flavours. In particular, in the lepton sector, the same
mass insertions which induce EDMs are sources of lepton flavour violation,
again via neutralino or chargino loop diagrams~\cite{Hisano:1995cp}. As a consequence, we expect a
correlation between EDM and LFV processes and the allowed parameter space to
be strongly constrained by the experimental limits on LFV decays such as
$BR(l_i \to l_j\; \gamma)$.

 The branching ratio of $l_{i}\rightarrow l_{j}\gamma$
 can be written as
 \bea
 \frac{BR(l_{i}\rightarrow  l_{j}\gamma)}{BR(l_{i}\rightarrow
l_{j}\nu_i\bar{\nu_j})} =
 \frac{48\pi^{3}\alpha}{G_{F}^{2}}(|A_L^{ij}|^2+|A_R^{ij}|^2)\,,
\eea
 with the SUSY contribution to each amplitude given by the sum of two terms
$A_{L,R}=A_{L,R}^{n}+A_{L,R}^{c}$, where $A_{L,R}^{n}$ and $A_{L,R}^{c}$ denote the
contributions from the neutralino and chargino loops respectively.
 In the mass insertion approximation, and taking only the dominant terms,
we can write the amplitudes as follows:
 \bea
 \label{MIamplL}
 A^{ij}_{L}&=&\Frac{\alpha_{2}}{4\pi} \frac{
   (\delta^e_{\LL})_{ij} }{m_{\tilde l}^{2}}
 ~\Bigg[
 \frac{\mu M_{2}\tan\beta}{(M_{2}^2\!-\!\mu^2)}~
 F_{2\LL}(a_2,b)\!+\! \tan^2\theta_{W}\,  \frac{\mu
   M_{1}\tan\beta}{(M_{1}^2\!-\!\mu^2)}~F_{1\LL}(a_1,b)~\Bigg]\\ 
 &+& \Frac{\alpha_{1}}{4\pi}~
 \frac{(\delta^e_{\RL})_{ij}}{m_{\tilde
l}^{2}}~\left(\frac{M_1}{m_{l_i}}\right)~F_{1\LR}(a_1)\,,\nonumber
 \eea
 \bea
 \label{MIamplR}
 A^{ij}_R=\frac{\alpha_{1}}{4\pi}&\!\!\Bigg(\!\!&\frac{ (\delta^e_{\RR})_{ij}}{m_{\tilde
l}^{2}} \frac{\mu M_{1}\tan\beta}{(M_{1}^2\!-\!\mu^2)}~ F_{1\RR}(a_1,b)\!+\!
 \frac{(\delta^e_{\LR})_{ij}}{m_{\tilde l}^{2}}~\left(\frac{M_1}{m_{l_i}}\right)~
F_{1\LR}(a_1) \Bigg)\,, 
 \eea
where $\theta_W$ is the weak mixing angle, $a_{1,2}=M^{2}_{1,2}/\tilde{m}^{2}$,
$b=\mu^2/m_{\tilde l}^{2}$ and the loop functions
$F_{1\LL}$, $F_{2\LL}$, $F_{1\RR}$ and $F_{1\LR}$ can be obtained from the
expressions in Refs.~\cite{Paradisi:2005fk,Ciuchini:2007ha}.

Here we can see that $(\delta^e_{\LL})_{ij}$ and $(\delta^e_{\RR})_{ij}$
contributions are $\tan\beta$-enhanced. In contrast to MFV models with RH
neutrinos, in our model the largest contribution to $\mu\to e\gamma$ comes
from the RR sector, being $(\delta^e_{\RR})_{12}\simeq \bar\varepsilon^3$
while $(\delta^e_{\LL})_{12}\simeq \varepsilon^2\bar\varepsilon$ (see
Section~\ref{sec:flavmodel}). The $\tau\to\mu\gamma$ and $\tau\to e\gamma$
decays shall have similar LL and RR contributions. 

On the other hand, the only term proportional to $(\delta^e_{\LR})_{ij}$ 
arises from pure $\tilde B$ exchange and it is completely
independent of $\tan\beta$. However, in these flavour models LR mass 
insertions can be still important. In the case of the $\mu \to e \gamma$ decay
and taking into account the necessary chirality change in the amplitude,  we
have to compare $(\delta^e_{\RR})_{12}~m_{\mu} \tan\beta$ with
$(\delta^e_{\LR})_{12}~M_1$, as we can see from \eq{MIamplR}. Using the
expression 
for the mass insertions of \eq{delta23}, we see that in these models: 

\bea  
\Frac{(\delta^e_{\RR})_{12}~ m_{l_i} \tan
   \beta}{(\delta^e_{\LR})_{12}~ M_1} \simeq \Frac{(\bar
  \varepsilon^3/3)~ m_{\mu} \tan \beta}{\bar \varepsilon^3 A_0~
  ( m_\tau/M_{\tilde e}^2)~ M_1} = \frac{m_{\mu}\tan \beta}{3 ~ m_\tau}
\frac{M_{\tilde e}^2}{A_0 M_1}\, .  
 \eea 

 Therefore, we can see that if $M_{\tilde e}^2/(A_0 M_1)\sim O(1)$, the LR mass
insertion will dominate the $\mu \to e \gamma$ decay up to $\tan \beta \sim
 30$. In fact, these contributions can easily bring  $BR(\mu\to e \,\gamma)$ 
to the level of the present experimental reach and therefore, we expect that 
 the $A_0\neq 0$ scenarios will be very
strongly constrained by the present and future limits on $BR(\mu\to e
\,\gamma)$.
This is the main consequence of the misalignment between $A_e$ and
$Y_e$.  Let us notice that here the LR contribution, even if not enhanced by
$\tan\beta$, becomes dominant due to an enhancement by a factor of order 
$m_\tau/m_\mu$ with respect to the other contributions to the amplitude.
This is clearly peculiar of $\mu \to e \gamma$ and it is not verified in the case of
$\tau \to \mu \gamma$. 
For $\tau\to \mu \,\gamma$, even with $A_0 \neq 0$ the LR contribution is 
subdominant with respect to the other ones, 
mainly proportional to $(\delta^e_{\LL})_{32}$, which are $\tan\beta$-enhanced. 
Therefore, we expect the $BR(\tau\to \mu\,\gamma)$ for $A_0 \neq 0$ 
to be approximately equal to the case $A_0 = 0$. 

It is also interesting to compare the different LFV channels.
In the case $A_0=0$, the dominant LFV source should be $\delta^e_{\LL}$ which
contributes both to chargino and neutralino diagrams.  Thus a rough estimation
for the relative sizes of the branching ratios can be:  
\begin{eqnarray}
\label{brs1}
{BR(\tau\to e \,\gamma) \over BR(\mu\to e\,\gamma) } &  \approx &
\left({m_\tau \over m_\mu}\right)^5  {\Gamma_\mu \over \Gamma_\tau} 
{(\delta^e_{\LL})^2_{13} \over  (\delta^e_{\LL})^2_{12}} \approx 
{\mathcal O}(1) \\
\label{brs2}
{BR(\tau\to\mu\,\gamma) \over BR(\mu\to e\,\gamma) } & \approx &
\left({m_\tau \over m_\mu}\right)^5  {\Gamma_\mu \over \Gamma_\tau} 
{(\delta^e_{\LL})^2_{23} \over  (\delta^e_{\LL})^2_{12}} \approx 
{\mathcal O}(10^3)  
\end{eqnarray}
where $\Gamma_\mu$ ($\Gamma_\tau$) is the $\mu$ ($\tau$) full width. Given the
present limit $BR(\tau\to e \,\gamma) < 1.1 \times 10^{-7}$
\cite{Aubert:2005wa}, we can see from Eq.~(\ref{brs1}) that $\tau\to e
\,\gamma$ is not able to constrain the parameter space of the model better
than $\mu\to e\,\gamma$ whose experimental bound is $BR(\mu\to e \,\gamma) <
1.2 \times 10^{-11}$ \cite{Ahmed:2001eh} (which will be improved by two orders
of magnitude by MEG \cite{meg}). On the other hand, we expect from
Eq.~(\ref{brs2}) that the present constraints given by $\mu\to e \,\gamma$ and
$\tau\to \mu \,\gamma$ are comparable, once the combined BaBar+Belle limit
$BR(\tau\to \mu \,\gamma) < 1.6 \times 10^{-8}$ is considered
\cite{Banerjee:2007rj, Lusiani:2007cb}.

It is important also to clarify the dependence of our results on the chosen
value of $Y_{\nu}$. Notice that, in our $SU(3)$ model, the value of the
neutrino Yukawa couplings are fixed by the symmetry following
\cite{Ross:2004qn}. However, different values of $Y_{\nu}$ could be possible
in other examples while respecting to observed values of neutrino masses and
mixings. In any case, as can be seen in Eqs.~(\ref{soft2}), only the $(1,3)$
and $(2,3)$ elements of $m^2_{\tilde{L}}$ depend on the value of
$y^{\nu}_{33}$ at 1 loop. Therefore only the predictions on $\tau\to l_i
\,\gamma$ can be affected by a change on $y^{\nu}_{33}$ and even in this case
the contributions from $m^2_{\tilde{e}}$, independent on $y^{\nu}_{33}$, will
be of a similar size.

\section{Numerical results}
\label{sec:numerical}
In the following, we shall use the expressions for $d_e$ and LFV processes 
to put bounds on the SUGRA parameter space through a scan in $m_0$ and 
$M_{1/2}$ for fixed values of $\tan\beta$ and $a_0 = A_0/m_0$. Since 
all of our predictions will depend on arbitray $O(1)$ coefficients, it is not
possible to provide a precise numerical result. Thus, the following discussion
intends to point out the expected order of magnitude for these observables,
and factors of 2 or even 4 can appear. 

Our numerical analysis presented below is done defining the Yukawa, trilinear
and soft mass matrices at $M_{flav}=2\times10^{16}$ GeV, as explained in
Section~\ref{sec:flavmodel} and explicitely shown in the Appendix. 
Then we evolve the different flavour matrices to the
electroweak scale using 1-loop RGEs~\cite{Petcov:2003zb}. $O(1)$ coefficients in the
Yukawas matrices are determined by requiring a good fit on the fermion masses 
and quark mixings at $M_Z$~\cite{Roberts:2001}. Other $O(1)$ terms in the soft
matrices are taken as random, varying between 0.5 and 2. The different values
of $\tan\beta$ are fixed by the ratio of the vevs $(a^u_3/a^d_3)$, that in our
model are global factors in the Yukawa (and trilinear) matrices.

After running the resulting structures down to the $M_Z$ scale, we diagonalize
the Yukawas in order to obtain the left and right mixing matrices and rotate
the soft matrices into the SCKM basis. Notice that this SCKM rotation generates the
off-diagonal elements in the first row and column in all soft mass matrices and
in the case of left-handed sleptons generates also the dominant contribution to
$(M_{\tilde L}^2)_{23}$. %We assume that the lepton Yukawas have CKM-like mixings~\cite{Masiero:2002jn}, and that the observed neutrino mixing angles are due to the combination with the structure of the Majorana neutrino mass matrix.
At this scale, we apply  the LEP bounds on the lightest sparticle and Higgs
masses and we require a neutral LSP. In the plots, regions that fail to
satisfy any of these requirements are shown in dark brown (black). 
Then, we also apply the constraints set by the current bounds from $d_e$, LFV
processes and FCNC measurements on the
hadronic sector: $\Delta m_K$, $\epsilon_K$, $\Delta m_D$, $\Delta m_B$,
$\Delta m_{B_s}$ and $b\to s\gamma$~\cite{Yao:2006px}, as shown in
Table~\ref{Table:Constraints}. Nonetheless, only $\mu\to e\gamma$,
$\tau\to\mu\gamma$ and $\epsilon_K$ shall exclude regions in the parameter
space above the LEP and LSP bounds, we show these regions, stretching above
the dark brown regions at low $m_0$, in green (grey).

\begin{table}
\begin{center}
\begin{tabular}{|c|c||c|c|}
\hline
Observable & Bound & Observable & Bound \\
\hline \hline
$d_e$ & $<1.4\times10^{-27}$ e cm & $\Delta m_K$ & $<3.48\times10^{-15}$ GeV\\
$BR(\mu\to e\gamma)$ & $<1.2\times10^{-11}$ & $\Delta m_{D}$ & $<4.61\times10^{-14}$ GeV \\
$BR(\tau\to\mu\gamma)$ & $<4.5\times10^{-8}$ & $\Delta m_B$ & $<3.34\times10^{-13}$ GeV \\
$BR(\tau\to e\gamma)$ & $<1.1\times10^{-7}$ & $\Delta m_{B_s}$ & $<1.17\times10^{-11}$ GeV \\
$BR(b\to s\gamma)^{\rm SUSY}$ & $<0.88\times10^{-4}$ & $\epsilon_K$ & $<2.239\times10^{-3}$ \\
\hline
\end{tabular}
\end{center}
\caption{\label{Table:Constraints} Applied constraints coming from EDMs and LFV (left) and neutral meson sector (right).}
\end{table}

Initially, we present the results in a generic $SU(3)$ model with phases
$O(1)$ in the flavour soft terms as shown in the Appendix, \eq{soft0}. Then, we
take the RRV model as an explicit example with a well-defined phase structure
as shown in \eq{soft2}\footnote{In this model, even though each term in the
  soft matrices receives corrections from the RGEs, the leading terms from these corrections have an identical phase structure, so we can expect the phases not to change much by the running.}.

\subsection{Generic $SU(3)$ Model}

In this model, we assume generic $O(1)$ phases in the soft mass matrices at the
flavour-breaking scale as specified in \eq{soft0} of Appendix A. In this way,
we try to include generic models with different number of flavons or different 
contributions to the soft mass matrices.% as explained in section~\ref{sec:flavmodel}. %In consequence, the phases at the electroweak scale in the SCKM basis are still $O(1)$. 

As concluded on Section~\ref{sec:EDMsection}, the most important contributions
to $d_e$ come first from the
$(\delta^e_{\LL})_{13}(\delta^e_{\LR})_{33}(\delta^e_{\RR})_{31}$ insertion, and
then from $(\delta^e_{\LR})_{1i}(\delta^e_{\RR})_{i1}$. How significant is the
latter depends on the value of $A_0$ and $\tan\beta$. % Since the presence of a large phase on $(\delta^e_{\RR})_{31}$ would contribute through both terms, %while a phase on $(\delta^e_{\LL})_{13}$ or $(\delta^e_{\LR})_{13}$ would only contribute through one of these combinations.
In the numerical analysis we assume one $O(1)$ phase on each
rephasing-invariant combination and thus, here it is enough to put it on
$(\delta^e_{\RR})_{31}$.  This allows us to estimate
the largest area in the $m_0 - M_{1/2}$ plane into which EDM experiments could
probe.

\begin{figure}
\includegraphics[scale=.475]{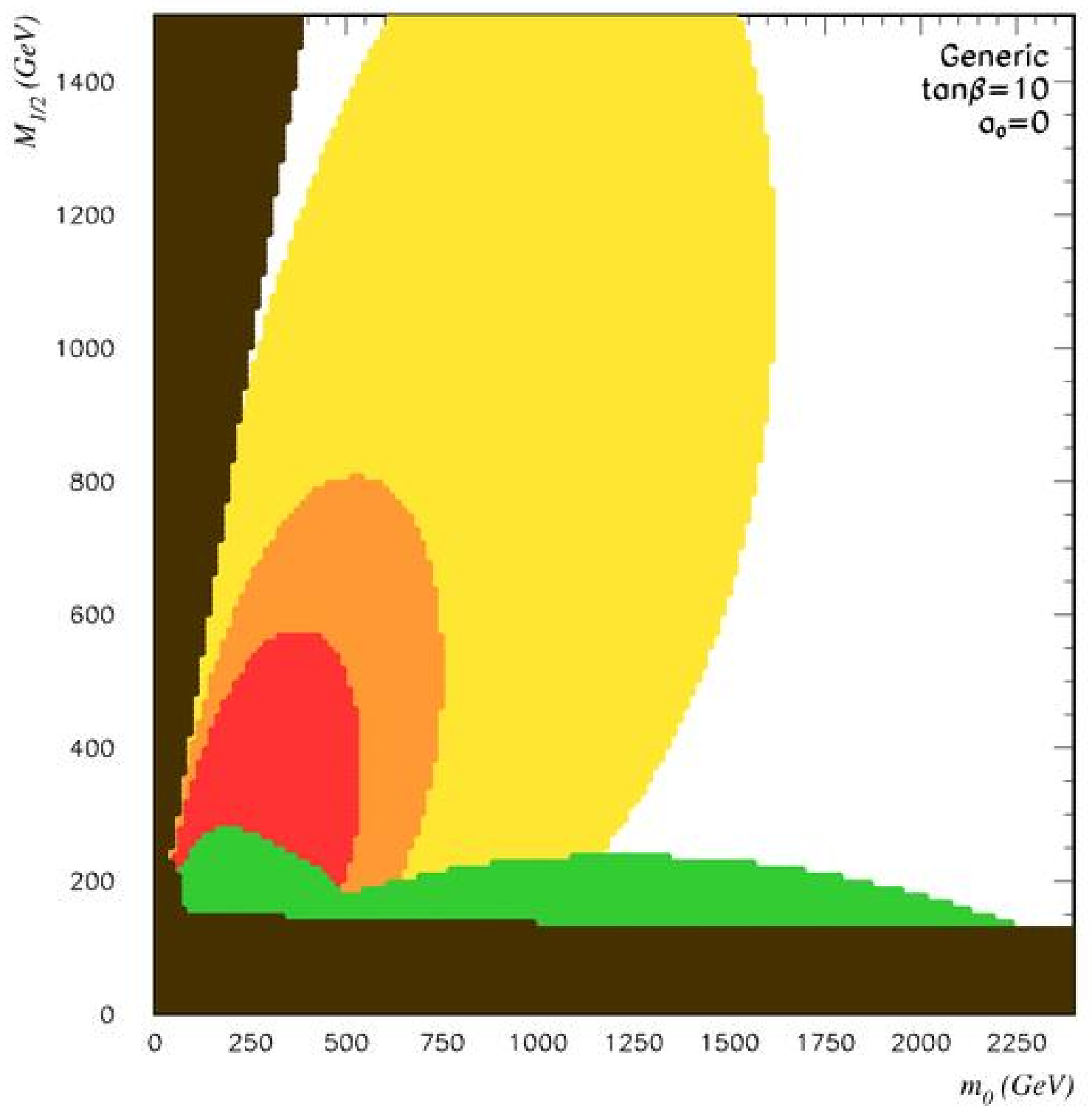} \includegraphics[scale=.475]{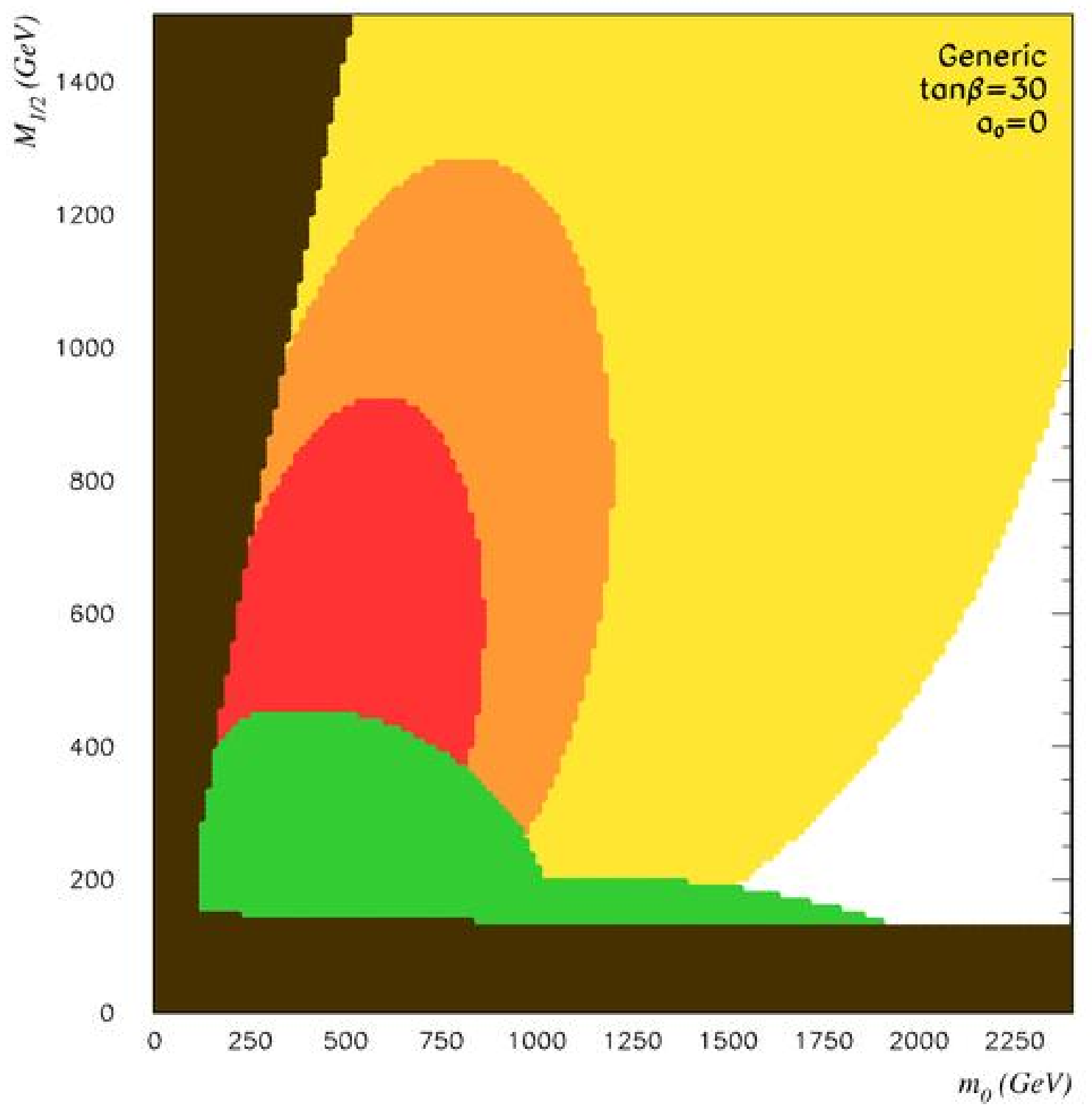}
\caption{Contours of $|d_e|=1\times10^{-28}$ e cm (red/dark grey),
  $|d_e|=5\times10^{-29}$ e cm (orange/medium grey) and
  $|d_e|=1\times10^{-29}$ e cm (yellow/light grey) in the $m_0$-$M_{1/2}$
  plane for $\tan\beta =10,30$ and $A_0=0$.  Current FCNC constraints and
  direct LEP bounds are also shown in green (grey) and dark brown (black)
  respectively.}
\label{fig:gen1}
\end{figure}

\begin{figure}
\includegraphics[scale=.475]{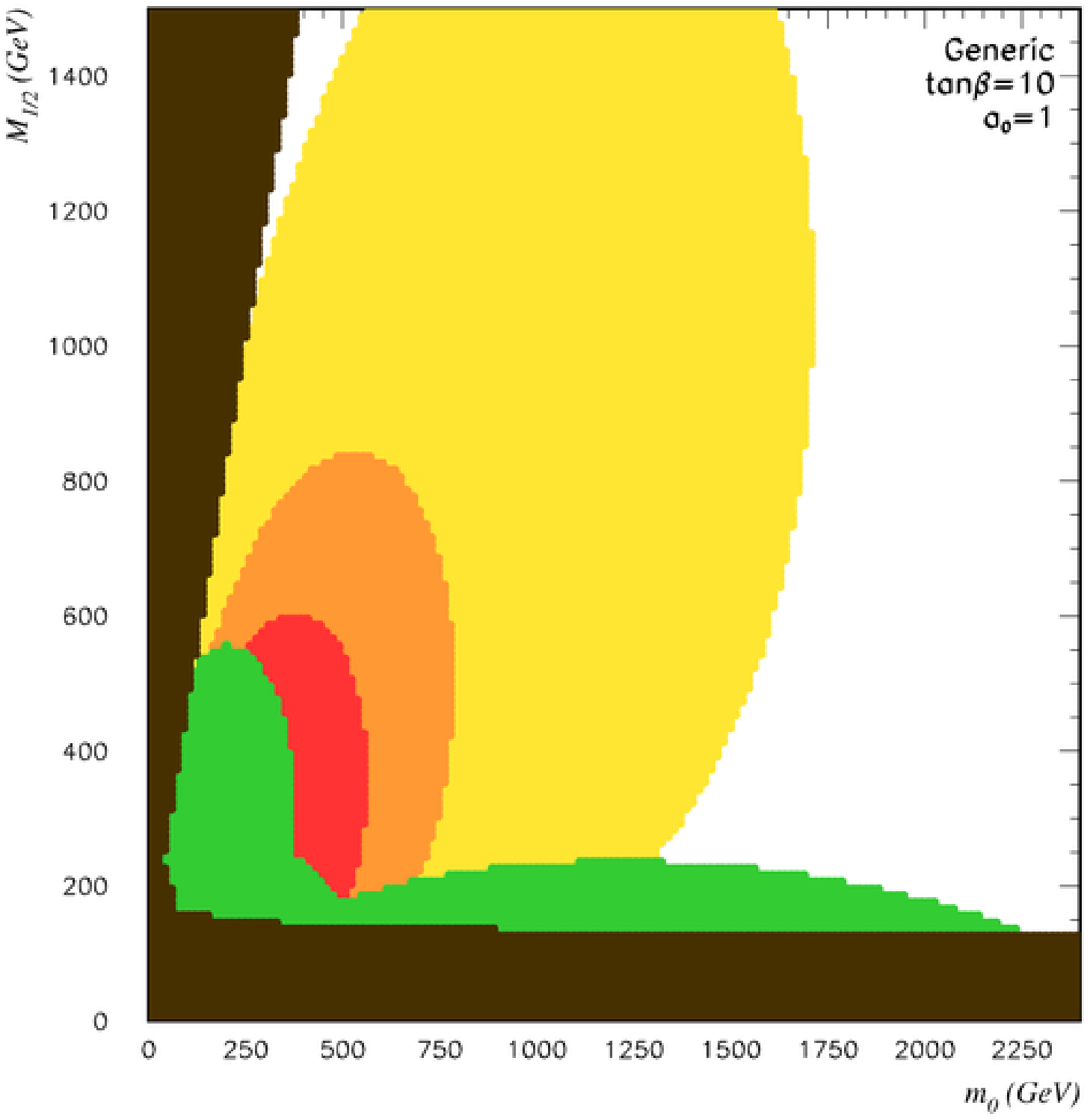} \includegraphics[scale=.475]{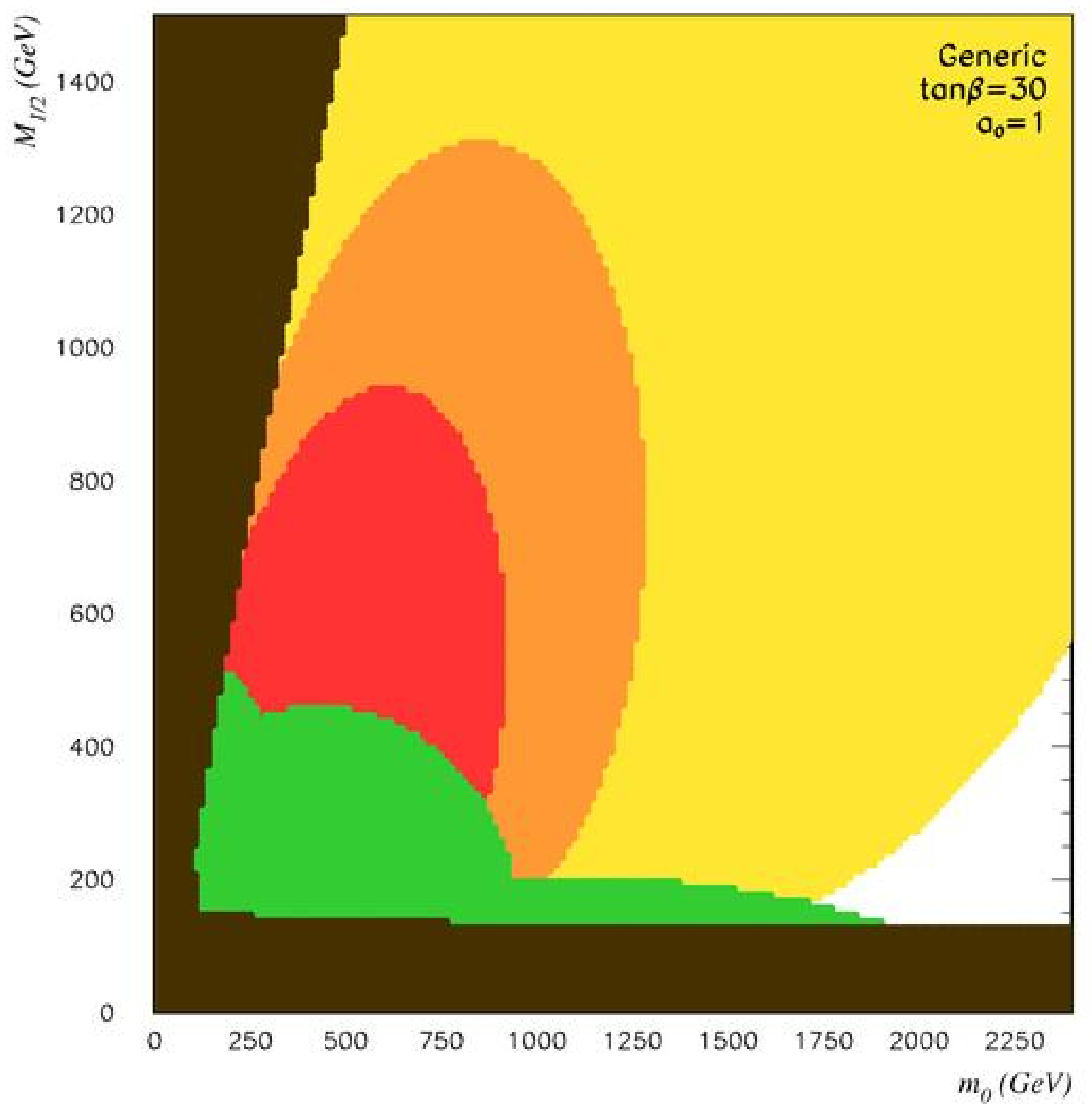}
\caption{Contours of $|d_e|$  and current constraints in the $m_0$-$M_{1/2}$
  plane for $\tan\beta =10,30$ and $a_0=1$. See caption of
  Figure~\ref{fig:gen1} for the meaning of different regions.}
\label{fig:gen2}
\end{figure}

In Figures \ref{fig:gen1} and \ref{fig:gen2} we show contours for expected
values of $|d_e|$ in the $m_0 - M_{1/2}$ plane. The red, orange and yellow
(dark grey, medium grey and light grey)
regions show contours for $|d_e|$ equal to $1\times10^{-28}$,
$5\times10^{-29}$ and $1\times10^{-29}$ e cm, respectively.

Figure~\ref{fig:gen1} takes $A_0=0$ and $\tan\beta=10,30$. In this case all off-diagonal
$\delta_{\LR}$ terms are generated by the running, and are thus small.
$d_e$ is then basically due to
$(\delta^e_{\LL})_{13}(\delta^e_{\LR})_{33}(\delta^e_{\RR})_{31}$ insertions.
It is important to emphasize that the present bound on $d_e$ does not provide a
constraint on the SUSY parameter space even for $\tan \beta=30$. However, by
reaching a sensitivity of $10^{-29}$ e~cm, we can explore values of $M_{1/2}$
and $m_0$ of the order of 1500 GeV. A value of $5\times 10^{-29}$ e~cm for
the electron EDM would explore the parameter space up to values of $M_{1/2}$
and $m_0$ of order 700 GeV. In other words, a reasonable value of the electron
EDM in these flavour models in the presence of large phases would be of the
order of $1.1 \times 10^{-28}$ e cm for $\tan \beta=10$ and  $3.0 \times 10^{-28}$
e cm for $\tan \beta=30$ with $(m_0,M_{1/2}) = (500, 300)$ GeV in both cases, 
corresponding
to an accessible sfermion spectrum at the LHC with squark masses around the 900
GeV.  
Thus, if large 
flavour phases are present and SUSY is to solve the SM hierarchy problem, we
can hope to find some signature of $d_e$ in the upcoming experiments, even for
low values of $\tan\beta$~\cite{WIP}.

In Figure~\ref{fig:gen2} we set $A_0=m_0$ ($a_0=1$). As expected, the
$(\delta^e_{\LR})_{13}(\delta^e_{\RR})_{31}$ insertion comes into play
especially for large $m_0$, due to the influence of the trilinears (remember
we take $A_0 = m_0$, i.e. it is not fixed to a single value), as they 
lower slepton masses in the RGEs. The expected values of $d_e$ are now similar 
to the values found in the $ A_0=0$ case, although can be slightly increased
in the regions of low $\tan \beta$ and large $m_0$ ($A_0$ for $a_0=1$). 
For comparison with the $A_0=0$ case, with 
 $(m_0,M_{1/2}) = (500, 300)$ GeV we obtain  $d_e \simeq 1.2 \times 10^{-28}$
 e cm for $\tan \beta=10$ and  $3.4 \times 10^{-28}$ e cm for $\tan \beta=30$.
Therefore, we see that in general they are slightly increased, although we
must keep in mind that having
a positive or negative interference  will depend on the phases. 
\begin{figure}
\includegraphics[scale=.475]{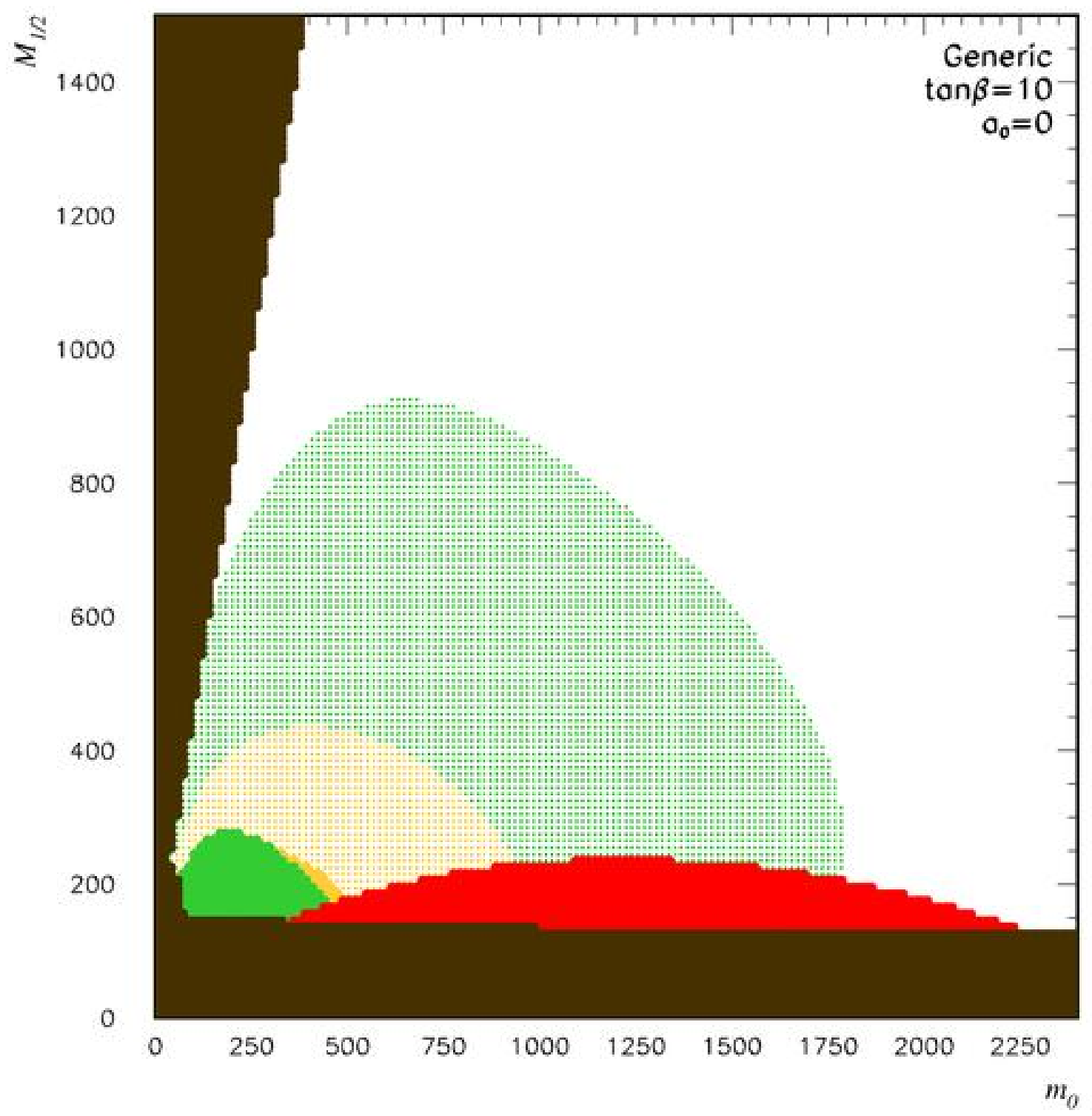} \includegraphics[scale=.475]{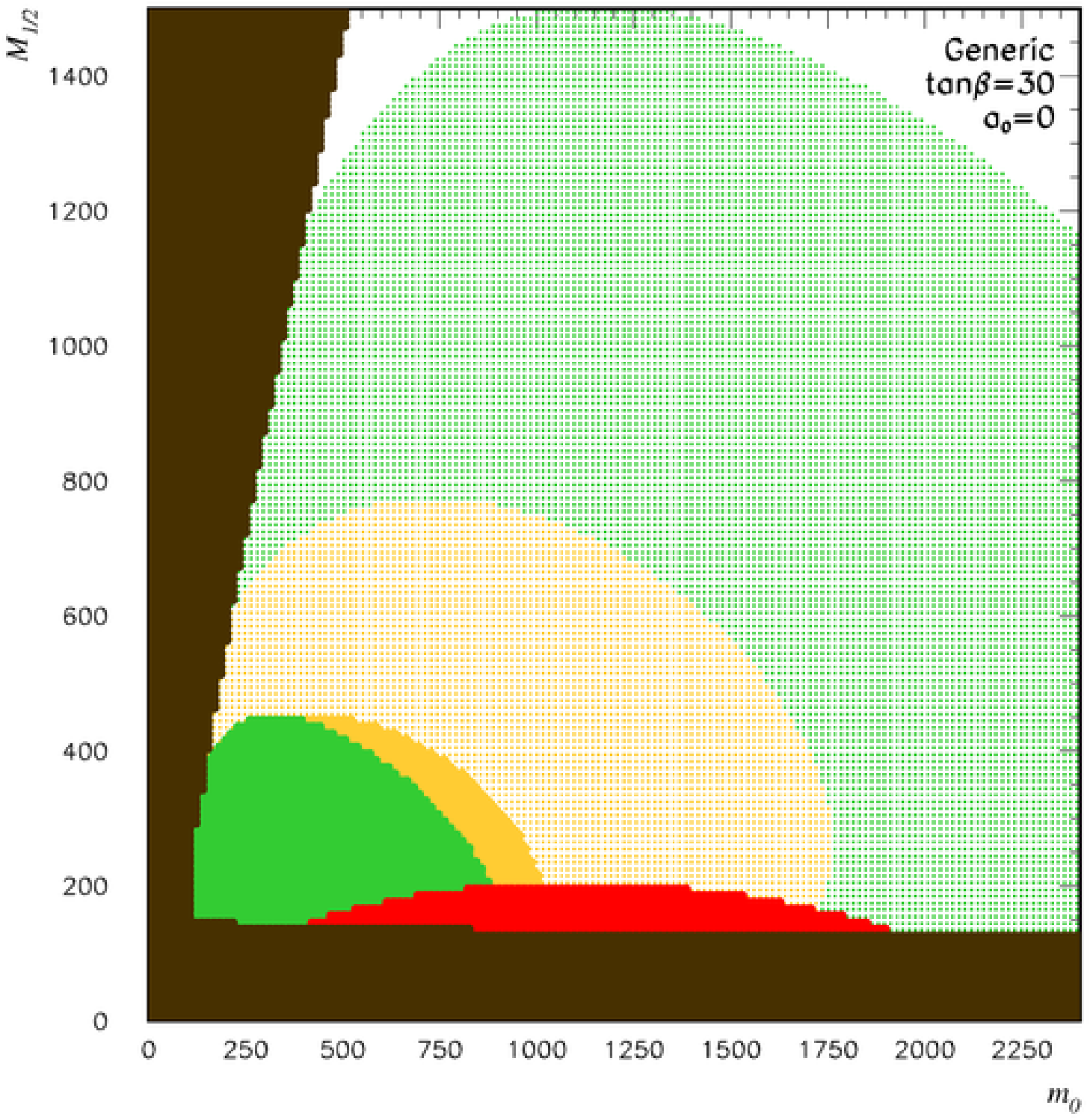}
\caption{Current constraints due to $\mu\to e\gamma$ (green/medium grey)),
  $\tau\to\mu\gamma$ (yellow/light grey) and $\epsilon_K$ (red/dark grey) in
  the $m_0$-$M_{1/2}$ plane for $\tan\beta =10,30$ and $A_0=0$. The green
  (medium grey) dotted region corresponds to the reach of $\mu\to e \gamma$ at
  the MEG experiment, while the yellow (light grey) hatched region is the reach of $\tau\to\mu\gamma$ at the Super Flavour Factory.}
\label{fig:gen1c}
\end{figure}

Figure~\ref{fig:gen1c} gives details on the current constraints given by LFV
experiments and $\epsilon_K$, which were shown previously in green (grey). 
We show the bounds of $\mu\to e\gamma$ (green/medium grey), $\tau\to\mu\gamma$
(yellow/light grey) and $\epsilon_K$ (red/dark grey). It is interesting to notice that, even though strong, the bounds still allow a very large area of the parameter space which is compatible with the observation of SUSY at the LHC.
The reach of the MEG experiment is also shown in Figure~\ref{fig:gen1c},
dotted in green (medium grey). We assume it capable of reaching a sensitivity
to the $\mu\to e \gamma$ branching ratio of $10^{-13}$~\cite{meg}. We also show the reach
of $\tau\to\mu\gamma$ experiments at the Super Flavour Factory, hatched in
yellow (light grey). We take the experiment to be able to measure the branching ratio down to $2\times10^{-9}$ \cite{Bona:2007qt}.

The impact of MEG in these $SU(3)$ flavour models on the evaluated parameter
space is impressive, covering values of $M_{1/2} \lsim 1500$ GeV and $m_0
\lsim 2500$ GeV for $\tan\beta=30$. Thus, if any evidence of SUSY is to be
found at the LHC, $\mu\to e\gamma$ decay should be seen at MEG. The same can
be said for $\tau\to\mu\gamma$ at the Super Flavour Factory, even though such constraints are not as strong for low $\tan\beta$.

\begin{figure}
\includegraphics[scale=.475]{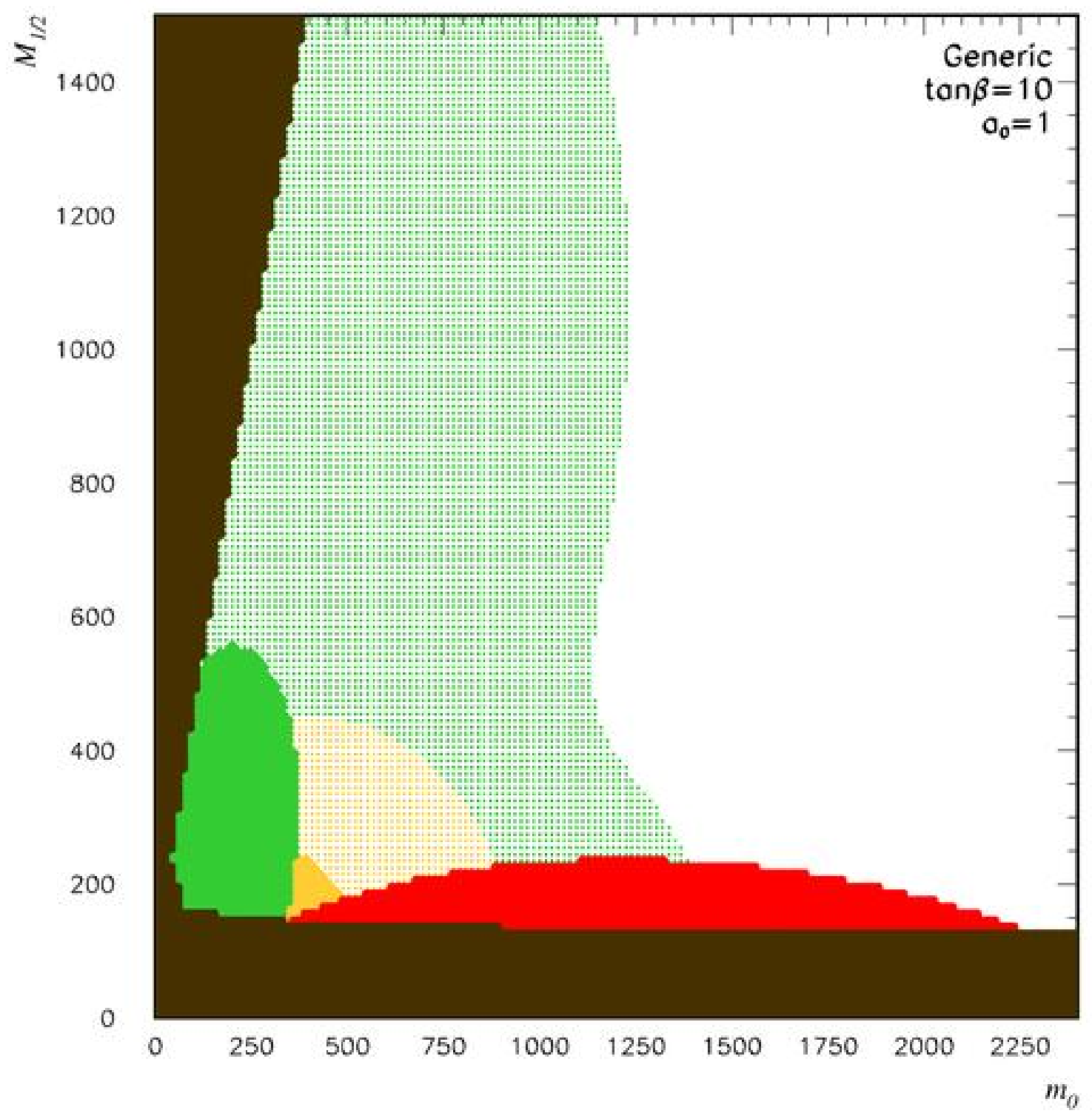} \includegraphics[scale=.475]{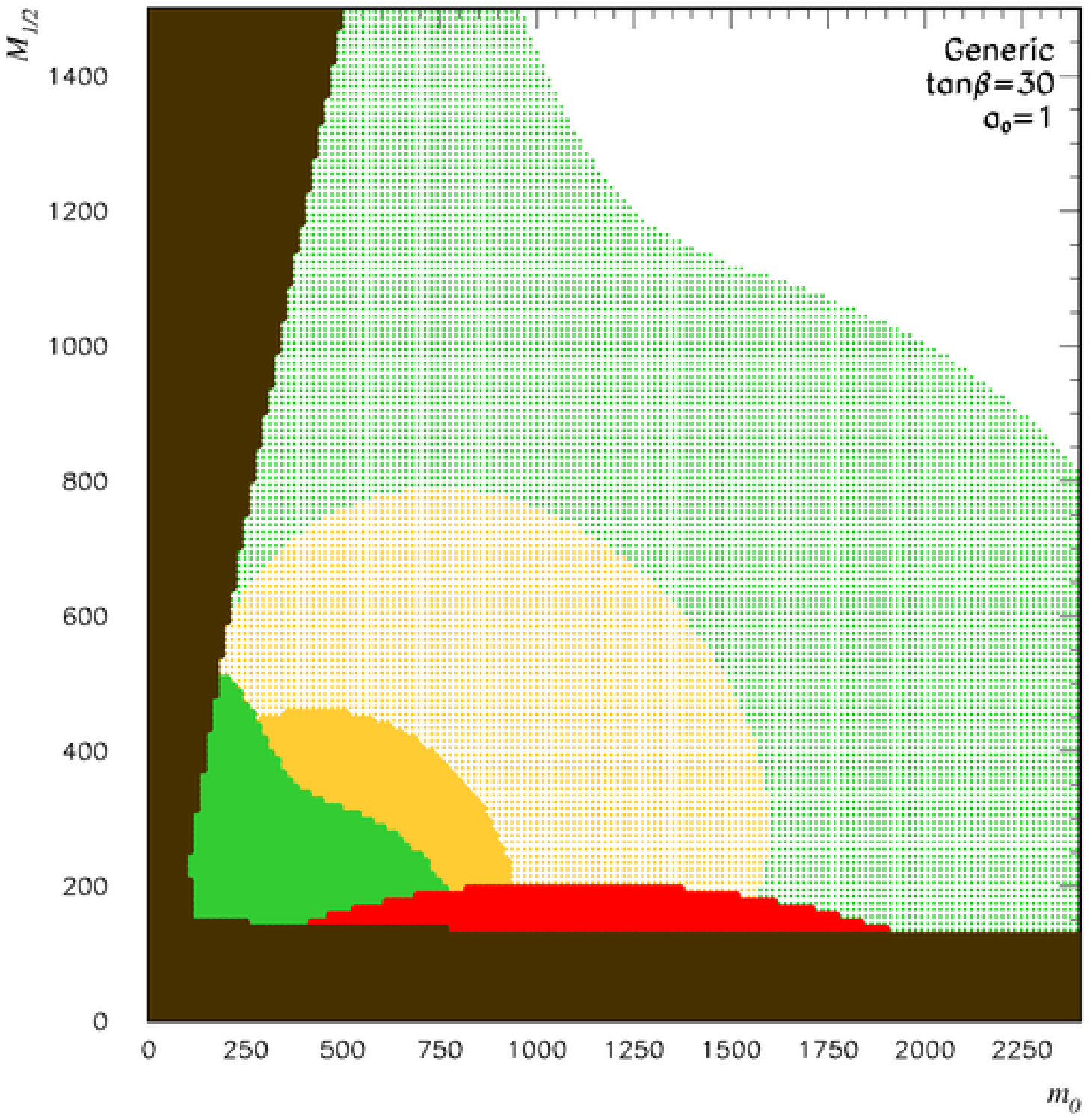}
\caption{Current constraints due to $\mu\to e\gamma$ (green/medium grey)),
  $\tau\to\mu\gamma$ (yellow/light grey) and $\epsilon_K$ (red/dark grey) in
  the $m_0$-$M_{1/2}$ plane for $\tan\beta =10,30$ and $a_0=1$. The green
  (medium grey) dotted region corresponds to the reach of $\mu\to e \gamma$ at
  the MEG experiment, while the yellow (light grey) hatched region is the
  reach of $\tau\to\mu\gamma$ at the Super Flavour Factory.} 
\label{fig:gen2c}
\end{figure}

The main effect of $A_0\neq0$ can be clearly seen in Figure~\ref{fig:gen2c}
in the decay $\mu\to e\gamma$. The appearance of a $(\delta^e_{\LR})_{12}$
term implies a considerable new neutralino contribution.
This new contribution can then interfere with the previous neutralino-chargino
diagrams. Positive or negative interference depends on the relative sign
(phase) between $(\delta^e_{\LR})_{12}$ and $(\delta^e_{\LL})_{12}$ as can be
seen in Eqs.~(\ref{MIamplL}) and (\ref{MIamplR}). From this Figure we can see
that this new contribution is indeed large and even dominant for $a_0=1$
specially for low values of $\tan \beta$. 

This new contribution is almost $\tan\beta$ independent, so we are allowed to
put strong bounds directly on the value of $A_0$. Notice that the constraints
coming from the \textit{current} bounds of $\mu\to e\gamma$ at small $m_0$ and
$M_{1/2}$ are already very strong. The MEG prediction, shown in Figure~\ref{fig:gen2c}, will now
cover the parameter space up to values of $M_{1/2} \lsim 1500$ GeV and $m_0
\lsim 1000$ GeV for $\tan\beta =10$ and $M_{1/2} \lsim 1500$ GeV and $m_0
\lsim 2500$ GeV for $\tan\beta =30$.

The $\tau\to\mu\gamma$ branching ratio is not affected as much by the flavour violating trilinear terms. The reason for this is that the other dominant insertions are proportional to $m_\tau\tan\beta$, as explained in Section~\ref{sec:MuegSection}. 
In contrast, $d_\mu$ and $d_\tau$, even though larger than $d_e$, can not be probed by the upcoming experiments. As explained in section
\ref{sec:EDMsection}, in these flavour models, the muon
EDM is typically two orders of magnitude
larger than the electron EDM. In this case this would imply that $d_\mu$ can
be at most of the order of $10^{-26}$ e cm, still orders of magnitude below
the reach of the proposed experiments.

\subsection{Generalization of the RVV model with spontaneous CP}

In this section, we analyze a variation of the model defined 
in~\cite{Ross:2004qn} and presented in the Appendix. 
As can be seen in \eq{soft2}, we have only one physical phase at leading order 
in the soft mass matrices in the lepton sector: $(\beta_3-\chi)$. Notice
that this is due to the fact that, in this model, the soft-breaking terms have
the ``minimal'' structure given by \eq{eq:minimal}. This can change if
different operators like  $\theta_{23,i} \overline{ \theta}_{3}^{j} + h.c.$
contribute to the soft mass matrices.  

\begin{figure}
\includegraphics[scale=.475]{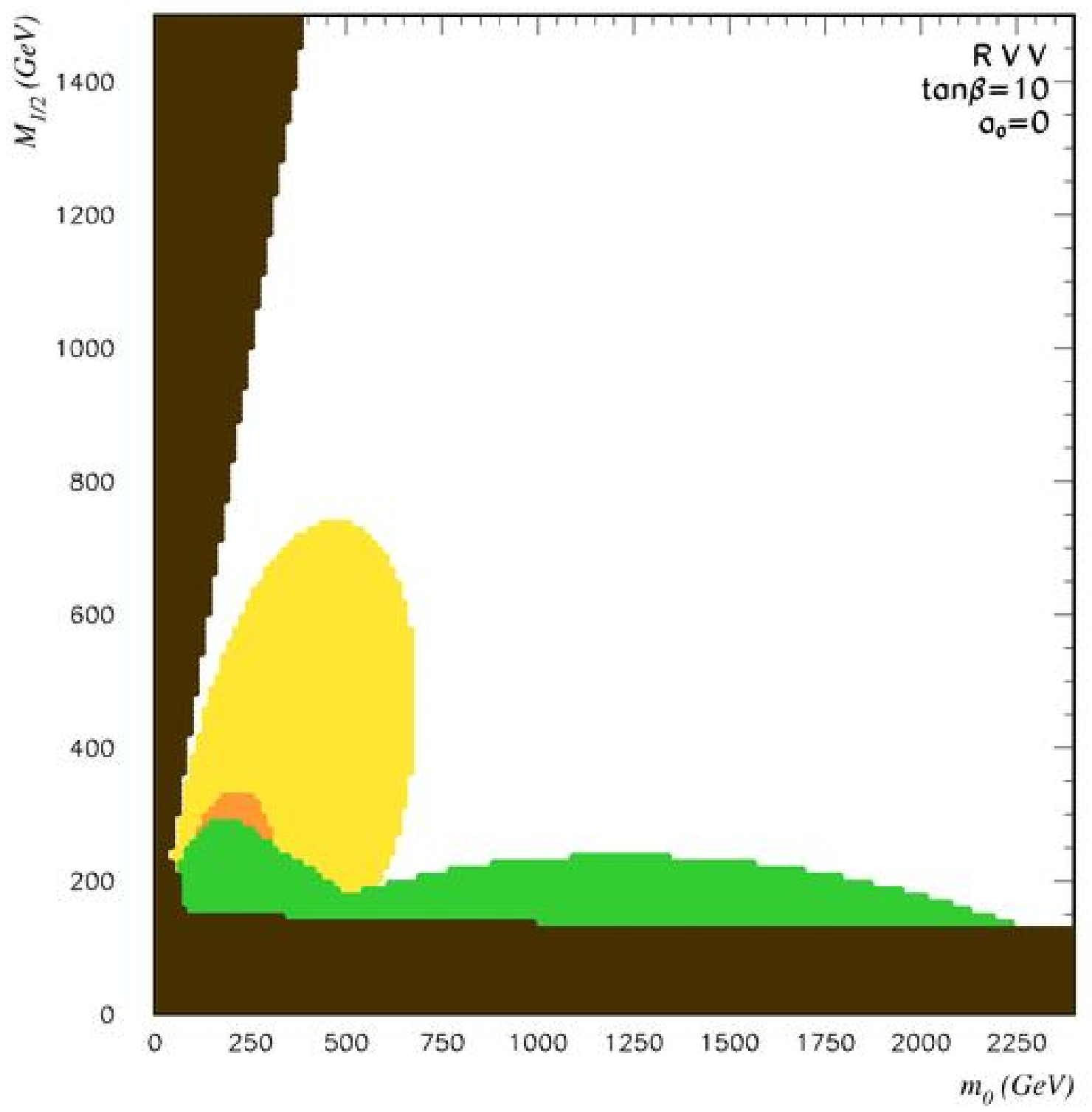} \includegraphics[scale=.475]{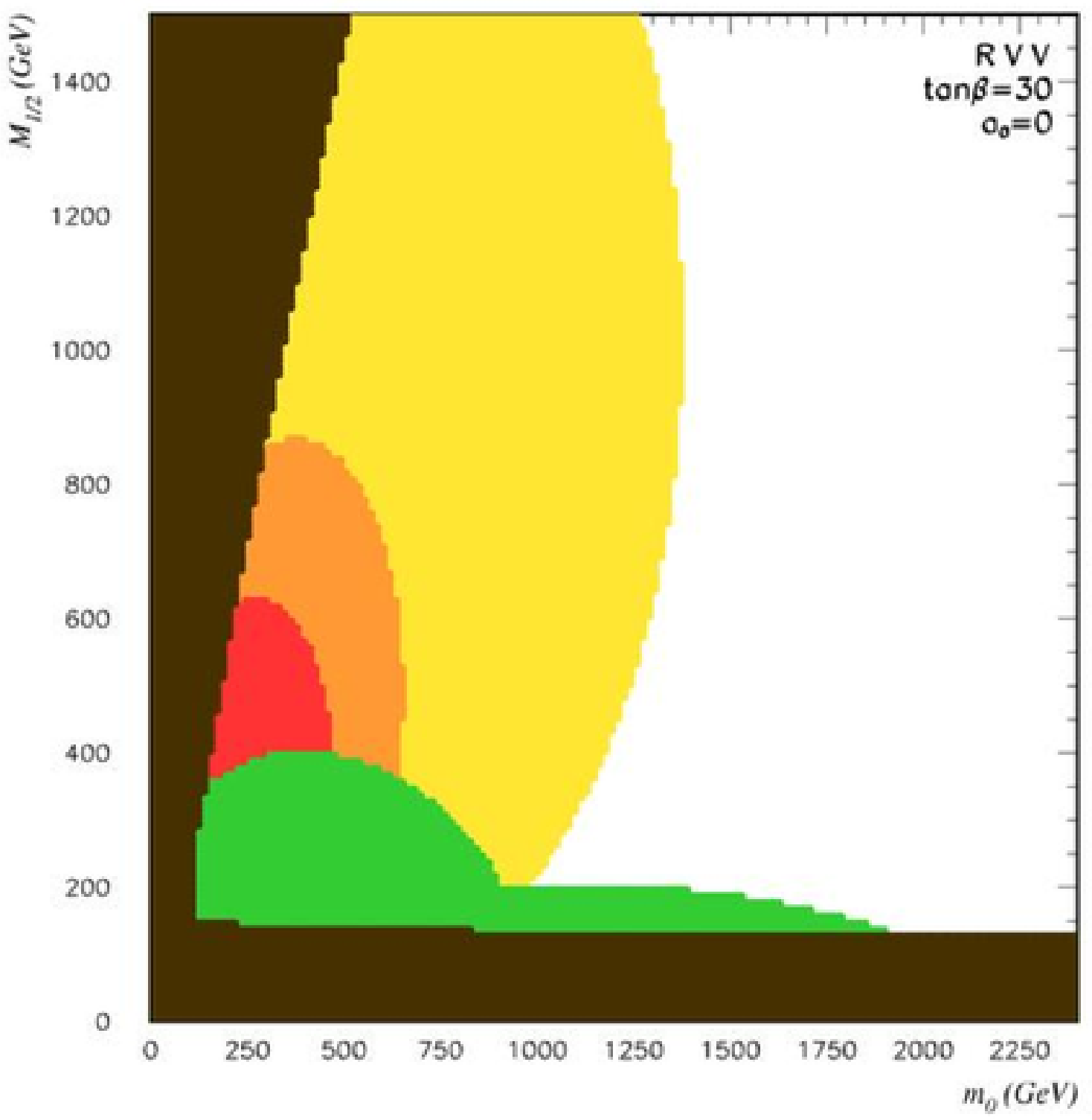}
\caption{Contours of $|d_e|$ as in Figure~\ref{fig:gen1}, for $A_0=0$, but with phases following the RVV model.}
\label{fig:mod1}
\end{figure}

With $A_0=0$, we can see that the leading phases in the
$(\delta^e_{\LL})_{13}$ and $(\delta^e_{\RR})_{31}$ are equal and therefore they
cancel in
$(\delta^e_{\LL})_{13}(\delta^e_{\LR})_{33}(\delta^e_{\RR})_{31}$\footnote{Notice
  that these phases are observable and will contribute to other CP violation
  observables \cite{Botella:2004ks}.}. The most significant contribution to $d_e$ 
shall depend on the phase of the subleading term $L_1$, $2(\beta_3-\chi)$. 
Due to this fact, and taking into account $\varepsilon/\bar \varepsilon =
1/3$, we can expect $d_e$ to be smaller than in the generic case roughly by a
factor of 10. This is confirmed by the numerical results shown in 
Figure~\ref{fig:mod1}.

\begin{figure}
\includegraphics[scale=.475]{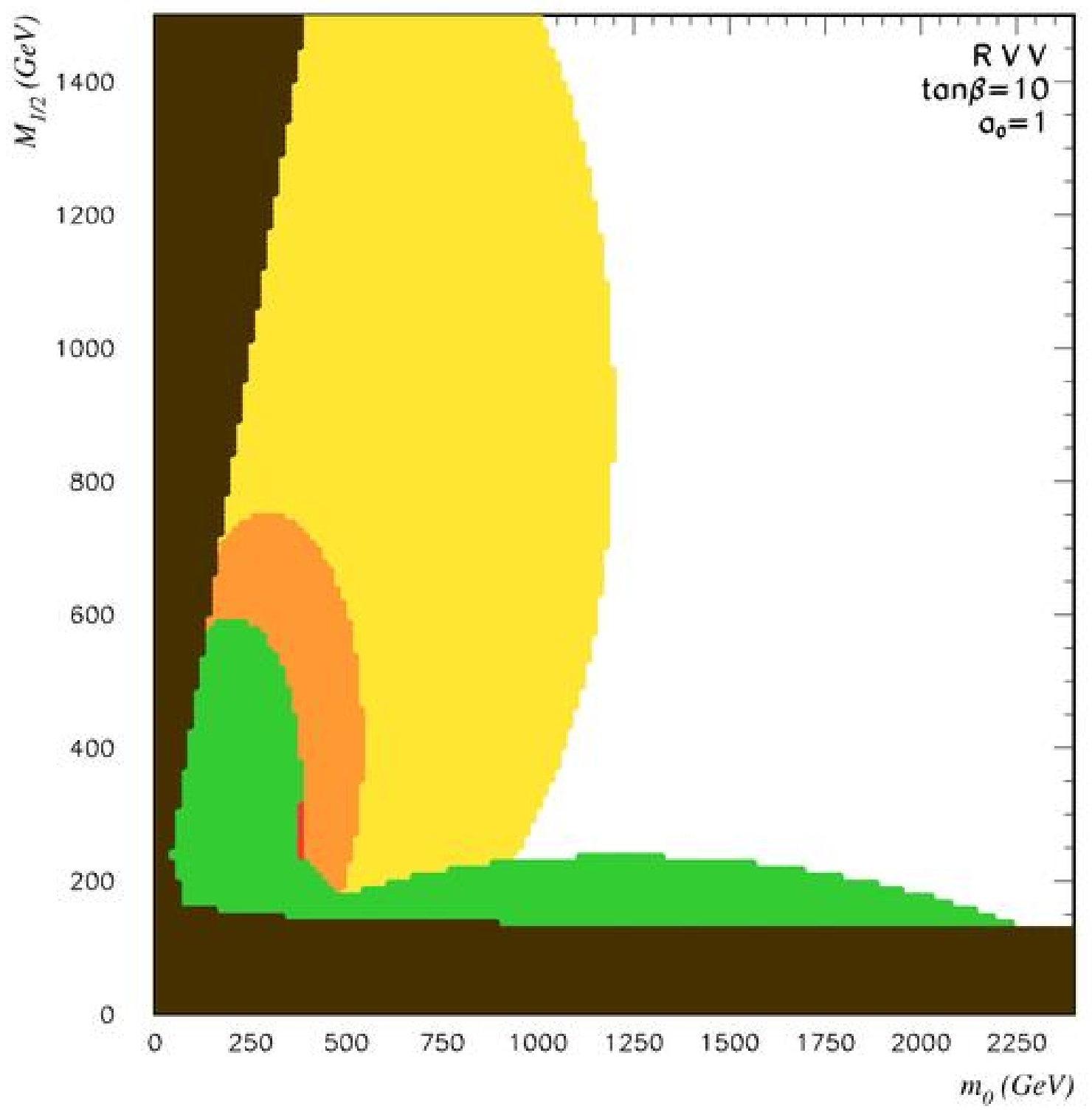} \includegraphics[scale=.475]{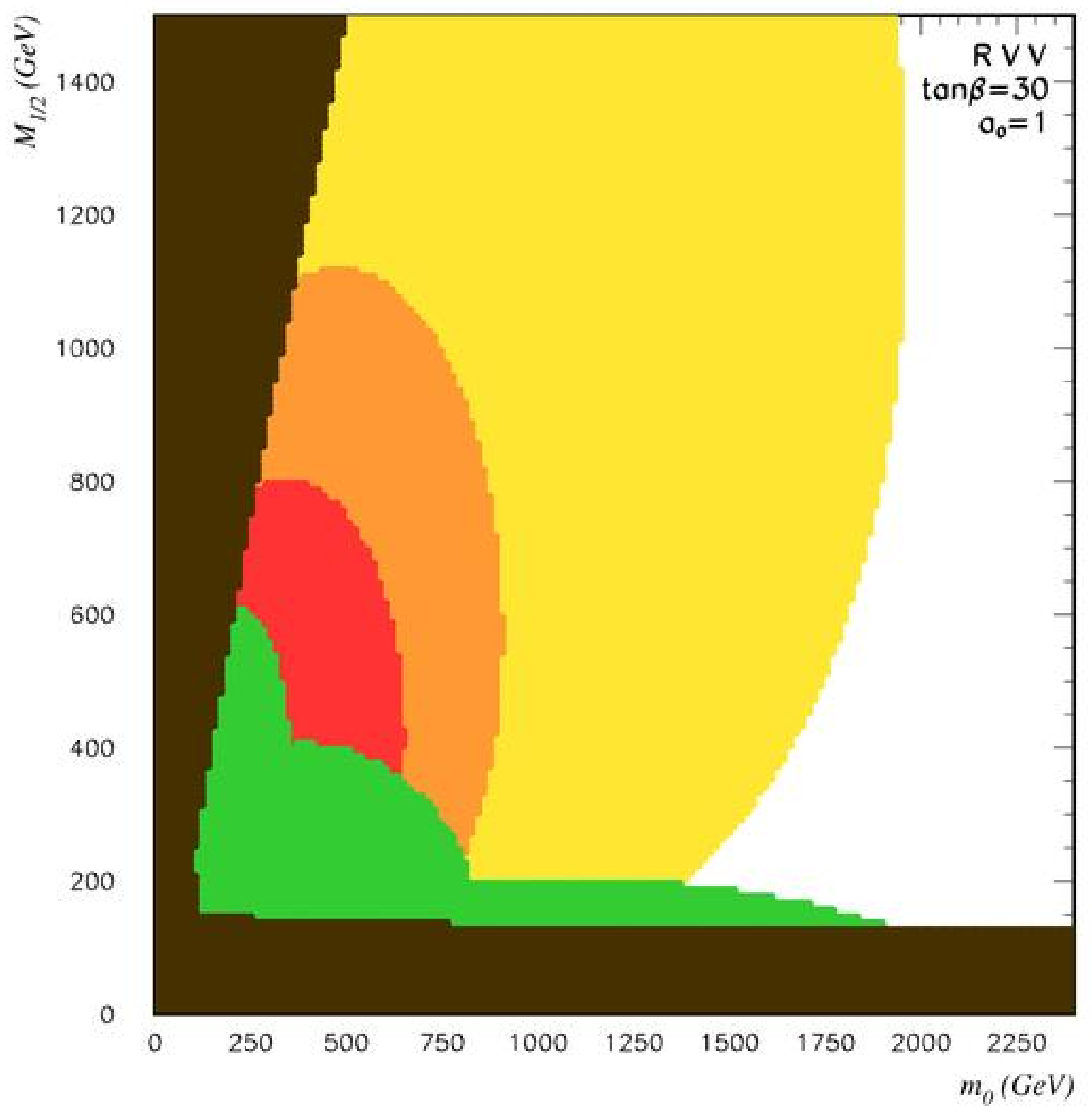}
\caption{Contours of $|d_e|$ as in Figure~\ref{fig:gen1}, for $A_0=m_0$, but with phases following the RVV model.}
\label{fig:mod2}
\end{figure}

On the other hand, for $A_0\neq0$, the phase $2(\beta_3-\chi)$ appears at 
leading order in the combination
$(\delta^e_{\LR})_{13}(\delta^e_{\RR})_{31}$. 
As we have seen in the previous section, with phases $O(1)$ in all MIs, 
this contribution is comparable to the
$(\delta^e_{\LL})_{13}(\delta^e_{\LR})_{33}(\delta^e_{\RR})_{31}$ contribution
in the large $m_0$ (and thus large $A_0$) region. Therefore, in this model, 
where $(\delta^e_{\LL})_{13}(\delta^e_{\LR})_{33}(\delta^e_{\RR})_{31}$ is reduced
by roughly an order of magnitude with respect to the generic case, 
we can expect the double mass insertion to dominate in most of the parameter
space. In fact, for small $\tan\beta$, the moduli of this double MI is of the 
same order of magnitude as the triple MI and so we can expect to reach similar 
$d_e$ values to those obtained in the generic case. Even for larger $\tan
\beta\simeq 30$ the double MI is comparable to the triple MI in the large 
$m_0$ region where again the results for $d_e$ are similar to those obtained 
in the generic case. Therefore, with $A_0\neq0$ we can expect similar values
for $d_e$ as in the generic model. This can be seen in Figure~\ref{fig:mod2}.

The discussion of the constraints coming from MEG and Super Flavour Factories 
are completely analogous to those of the generic model, since these LFV 
processes do not depend heavily on the presence of sizeable phases. Therefore 
Figs.~\ref{fig:gen1c} and \ref{fig:gen2c} remain valid also in this model.

In summary, in this generalization of the RVV model with fixed phases, we
would explore values of $M_{1/2}$ and $m_0$ of the order of 800 and 600 GeV
with a value of $1\times 10^{-29}$ e~cm for
the electron EDM with $\tan  \beta=10$ and $A_0=0$. This means, we would need
an increase of 10 in the sensitivity to $d_e$ to explore the same region of
parameter space as in the generic model. However, if $a_0=1$ we would explore
a region very similar to the region explored in the generic $SU(3)$ model,
with similar values of $M_{1/2}$ and values of $m_0$ roughly smaller by a 
factor of 2. 
Reasonable values of the electron EDM in this explicit example for
$(m_0,M_{1/2}) = (500, 300)$ GeV and $A_0=0$ would be of the order of 
$1.8 \times 10^{-29}$ e cm for $\tan \beta=10$ and $8.3 \times 10^{-29}$ e cm
for $\tan \beta=30$. Thus, we see $d_e$ is reduced roughly a factor of 5 with
respect to the generic model.  For the same scalar and gaugino masses and 
$A_0=m_0$, $d_e \simeq 5.6 \times 10^{-29}$ e cm for $\tan \beta=10$ and 
$d_e \simeq 1.7 \times 10^{-28}$ e cm for $\tan \beta=30$, i.e., only a factor
of two smaller than in the generic model.

\section{Conclusions}
We have shown that the flavour and CP problems in the supersymmetric 
extensions of the SM are deeply related to the origin of flavour (and CP) 
in the Yukawa matrices. It is natural to think that the same 
mechanism generating the flavour structures and giving rise to CP violation 
in the Yukawa couplings is responsible for the structure and phases in the
SUSY soft-breaking terms. A flavour symmetry with spontaneous CP violation
in the flavour sector can simultaneously solve both problems. 

In this paper, we have analyzed the phenomenology of a non-Abelian $SU(3)$ 
flavour symmetry. In this model, flavour-independent phases are naturally zero
and only flavour-dependent phases are present in the soft-breaking terms. 
We have studied the contributions to the leptonic electric dipole moments from
these flavour phases and we have shown that the future bounds on the electron
EDM will be able to explore a large part of the SUSY parameter space in these
models. Simultaneously we have analyzed the reach of the future MEG and Super
Flavour factories through Lepton Flavour Violation processes.
We have shown that we can expect signals of new physics both in EDM and LFV
experiments if the SUSY masses are accessible at the LHC.

\section*{Acknowledgments}

We acknowledges partial support from the Spanish MCYT FPA2005-01678. 
We thank the CERN theory group for hospitality at various stages of work. 
L.~C. acknowledges partial support from the foundation ``Angelo Della Riccia''
and the hospitality of the ``Departament de F\'{\i}sica Te\`orica''. 
J.~J.~P.~would like to give thanks to the University of Padua for its
hospitality during his stay, and to Paride Paradisi for useful discussions.

\appendix

\section{Soft Matrices in the SCKM Basis}

In this appendix we present the structure of the soft mass matrices in the
SCKM basis in generic flavour symmetry models with a symmetric texture in the
Yukawa matrices. We then present an explicit example based in the model of
Ref.~\cite{Ross:2004qn}.

In order to make a mass insertion analysis of a given process, one must have
all soft matrices at the SCKM basis  where Yukawa matrices are diagonal at the electroweak scale. However, we present here the structure of the soft matrices in the SCKM basis at the flavour scale and we will add the running effects later. For a generic model, assuming general phases (although taking into account that Yukawa and trilinear matrices are symmetric) and neglecting $O(1)$ constants, we have: 
\begin{subequations}
\label{soft0}
\begin{eqnarray}
 Y_e & = & \left(\begin{array}{ccc}
\frac{\bar\varepsilon^4}{3} & 0 & 0 \\
0 & 3\, \bar\varepsilon^2 & 0 \\
0 & 0 & 1
\end{array}\right)y_{33}^e \\
\frac{A_e}{A_0} & = & \left(\begin{array}{ccc}
\frac{\bar\varepsilon^4}{3} & \bar\varepsilon^3 ~e^{i \alpha_1} &
\bar\varepsilon^3 ~e^{i \beta_1} \\ 
\bar\varepsilon^3 ~e^{i \alpha_1}   & 3\,\bar\varepsilon^2 &
3\,\bar\varepsilon^2 ~e^{i \gamma_1} \\
\bar\varepsilon^3 ~e^{i \beta_1} & 3~\bar\varepsilon^2 ~e^{i \gamma_1} & 1
\end{array}\right) y_{33}^e\\
\frac{(m^2_{\tilde{e}_R})^T}{m_0^2} & = & \left(\begin{array}{ccc}
1+\bar\varepsilon^{2} y_{33}^e & \frac{1}{3}\bar\varepsilon^3~e^{i\alpha_2} &
\frac{1}{3}\bar\varepsilon^3~e^{i\beta_2} \\ 
\frac{1}{3}\bar\varepsilon^3~e^{-i\alpha_2} & 1+\bar\varepsilon^2 &
\bar\varepsilon^2 ~e^{i \gamma_2}\\ 
\frac{1}{3}\bar\varepsilon^3~e^{-i\beta_2} &\bar\varepsilon^2 ~e^{-i
  \gamma_2}& 1+ y_{33}^e 
\end{array}\right) \\
\frac{m^2_{\tilde{L}}}{m_0^2} & = & \left(\begin{array}{ccc}
1+\varepsilon^2 y_{33}^\nu & \frac{1}{3}\varepsilon^2\bar\varepsilon
~e^{i\alpha_3} & \bar\varepsilon^{3}~y_{33}^\nu ~e^{i\beta_3} \\ 
\frac{1}{3}\varepsilon^2\bar\varepsilon ~e^{-i\alpha_3} & 1+\varepsilon^2 &
\bar\varepsilon^{2}~ y_{33}^\nu ~e^{i \gamma_3} \\
\bar\varepsilon^{3}~y_{33}^\nu ~e^{-i\beta_3} & \bar\varepsilon^{2}~y_{33}^\nu ~e^{-i
  \gamma_3} & 1+y_{33}^\nu
\end{array}\right)
\end{eqnarray}
\end{subequations}
where $y_{33}^e = (\langle\theta_3^d\rangle/M_d)^2 = m_\tau/(v \cos \beta)$
and $y_{33}^\nu = (\langle\theta_3^u\rangle/M_d)^2 = m_t/(v \sin \beta)$.

\begin{table}
\begin{center}
\begin{tabular*}{1.00\textwidth}{@{\extracolsep{\fill}}|c|c c c c c c c c|}
\hline
${\bf Field}$ & $\psi$ & $\psi^c$ & $H$& $\Sigma$ & $\theta_3$ & $\theta_{23}$ & $\bar\theta_3$ &
$\bar\theta_{23}$~~~ \\
\hline
${\bf SU(3)}$ & 3 & 3 & 1 & 1 & $\bar3$ & $\bar3$ & 3 & 3~~~ \\
${\bf R}$ & 1 & 1 & 0 & 3 & -2 & -2 & -1 & 2~~~ \\
${\bf U(1)}$ & 1 & 1 & -2 & -1 & 0 & 1 & -1 & 0~~~ \\
${\bf Z_3}$ & 1 & 1 & 2 & 0 & 1 & 1 & -1 & 2~~~ \\
\hline
\end{tabular*}
\end{center}
\caption{\label{Table:Charges} Charges required to build satisfactory Yukawa matrices in the RVV Model.}
\end{table}

In the following we will present an explicit example of an $SU(3)$ flavour
symmetry model with spontaneous CP violation and the phase structure of the
different matrices completely determined. This model is a generalization of the RVV model of Ref.~\cite{Ross:2004qn} with arbitrary values of $\tan \beta$. The charges of the different flavon fields under the $SU(3)$ and global symmetries are shown in Table~\ref{Table:Charges}.

After spontaneous breaking of the flavour symmetry (and CP symmetry) the vevs
of the different fields are:
\begin{align}
\langle\theta_3\rangle=\left( 
\begin{array}{c}
0 \\ 
0 \\ 
1\end{array}
\right)&\otimes \left(
\begin{array}{cc}
a_3^u & 0 \\ 
0 & a_3^d~ e^{i \chi}
\end{array}
\right);& \langle\bar{\theta}_3\rangle=\left( 
\begin{array}{c}
0 \\ 
0 \\ 
1\end{array}
\right)&\otimes \left(
\begin{array}{cc}
a_3^u~ e^{i \alpha_u} & 0 \\ 
0 & a_3^d~ e^{i \alpha_d}
\end{array}
\right);  \nonumber \\
\langle\theta_{23}\rangle=&\left( 
\begin{array}{c}
0 \\ 
b_{23} \\ 
b_{23}~e^{i\beta_3}
\end{array}
\right);& \langle\bar\theta_{23}\rangle=&\left( 
\begin{array}{c}
0 \\ 
b_{23}~e^{i\beta^{\prime}_2} \\ 
b_{23}~e^{i(\beta^{\prime}_2 - \beta_3)}%
\end{array}
\right);
\end{align}
where we require the following relations:
\begin{align}
\label{expansion}
\left(\frac{\displaystyle{a_3^u}}{\displaystyle{M_u}}\right)^2& = y^{\nu}_{33},&
\left(\frac{\displaystyle{a_3^d}}{\displaystyle{M_d}}\right)^2& = y^e_{33}, \nonumber \\
\frac{\displaystyle{b_{23}}}{\displaystyle{M_u}}& = \varepsilon,& \frac{\displaystyle{b_{23}}}{\displaystyle{M_d}}& = \bar \varepsilon.
\end{align}

In the SCKM basis, and still neglecting $O(1)$ constants, the SUSY breaking matrices are:
\begin{subequations}
\label{soft2}
\begin{eqnarray}
\frac{A_e}{A_0} & = & \left(\begin{array}{ccc}
\frac{\bar\varepsilon^4}{3} & \bar\varepsilon^3 & \bar\varepsilon^3 ~e^{i(\beta_3-\chi)} \\
\bar\varepsilon^3 & 3\,\bar\varepsilon^2 & 3\,\bar\varepsilon^2 ~e^{i(\beta_3-\chi)}\\
\bar\varepsilon^3 ~e^{i(\beta_3-\chi)} & 3\,\bar\varepsilon^2 ~e^{i(\beta_3-\chi)} & 1
\end{array}\right)~y_{33}^e \\
\frac{(m^2_{\tilde{e}_R})^T}{m_0^2} & = & \left(\begin{array}{ccc}
1+\bar\varepsilon^{2}~y_{33}^e & \frac{1}{3}\bar\varepsilon^3 & \frac{1}{3}\bar\varepsilon^3~e^{-i(\beta_3-\chi)} \\
\frac{1}{3}\bar\varepsilon^3 & 1+\bar\varepsilon^2 & E^* ~\bar\varepsilon^2~e^{-i(\beta_3-\chi)} \\
\frac{1}{3}\bar\varepsilon^3~e^{i(\beta_3-\chi)} & E~\bar\varepsilon^2~e^{i(\beta_3-\chi)} & 1+y_{33}^e\end{array}\right) \\
\frac{m^2_{\tilde{L}}}{m_0^2} & = & \left(\begin{array}{ccc}
1+\varepsilon^2~y_{33}^\nu & \frac{1}{3}\varepsilon^2\bar\varepsilon & L_1^*~\bar\varepsilon^{3}~y_{33}^\nu ~e^{-i(\beta_3-\chi)} \\
\frac{1}{3}\varepsilon^2\bar\varepsilon & 1+\varepsilon^2 &
3 L_2^* ~\bar\varepsilon^{2}~y_{33}^\nu ~e^{-i(\beta_3-\chi)} \\
L_1~\bar\varepsilon^{3}~y_{33}^\nu ~e^{i(\beta_3-\chi)} &
3 L_2~\bar\varepsilon^{2}~y_{33}^\nu ~e^{i(\beta_3-\chi)} & 1+ y_{33}^\nu
\end{array}\right)
\end{eqnarray}
\end{subequations}
where the terms $E$, $L_1$ and $L_2$ include important subdominant
contributions with physical phases ($L_1$ and $L_2$ differ by $O(1)$ constants):
\begin{eqnarray}
 E & = & 1 - 3y_{33}^e~e^{-2i(\beta_3-\chi)} \\
 L_i & = & 1 - \frac{1}{3 y_{33}^\nu}\frac{\varepsilon^2}{\bar\varepsilon^2}
 e^{-2i(\beta_3-\chi)}\, .
\end{eqnarray}
In
order to reproduce down quark and electron masses for all values of
$\tan\beta$, we take $y_{33}^e=\langle\theta_3^d\rangle^2$ different from  $y_{33}^\nu=\langle\theta_3^u\rangle^2$.

Notice that, within this structure, the only physical phase is $\beta_3-\chi$. This is an additional phase with respect to the CKM phase $\omega$ in~\cite{Ross:2004qn}.


\begin{thebibliography}{0}
\expandafter\ifx\csname natexlab\endcsname\relax\def\natexlab#1{#1}\fi
\expandafter\ifx\csname bibnamefont\endcsname\relax
  \def\bibnamefont#1{#1}\fi
\expandafter\ifx\csname bibfnamefont\endcsname\relax
  \def\bibfnamefont#1{#1}\fi
\expandafter\ifx\csname citenamefont\endcsname\relax
  \def\citenamefont#1{#1}\fi
\expandafter\ifx\csname url\endcsname\relax
  \def\url#1{\texttt{#1}}\fi
\expandafter\ifx\csname urlprefix\endcsname\relax\def\urlprefix{URL }\fi
\providecommand{\bibinfo}[2]{#2}
\providecommand{\eprint}[2][]{\url{#2}}

\end{thebibliography}


\begin{thebibliography}{99}

%\cite{Masiero:2001ep}
\bibitem{Masiero:2001ep}
  A.~Masiero and O.~Vives,
  %``New physics in CP violation experiments,''
  Ann.\ Rev.\ Nucl.\ Part.\ Sci.\  {\bf 51} (2001) 161
  [arXiv:hep-ph/0104027].
  %%CITATION = ARNUA,51,161;%%

\bibitem{Masiero:2005ua}
  A.~Masiero, S.~K.~Vempati and O.~Vives,
  %``Flavour Physics and Grand Unification,''
  arXiv:0711.2903 [hep-ph].
  %%CITATION = ARXIV:0711.2903;%%


%\cite{Pospelov:2005pr}
\bibitem{Pospelov:2005pr}
  M.~Pospelov and A.~Ritz,
  %``Electric dipole moments as probes of new physics,''
  Annals Phys.\  {\bf 318} (2005) 119
  [arXiv:hep-ph/0504231].
  %%CITATION = APNYA,318,119;%%

%\cite{Raidal:2008jk}
\bibitem{Raidal:2008jk}
  M.~Raidal {\it et al.},
  %``Flavour physics of leptons and dipole moments,''
  arXiv:0801.1826 [hep-ph].
  %%CITATION = ARXIV:0801.1826;%%

%\cite{Ross:2000fn}
\bibitem{Ross:2000fn}
For a review and further references see:
\\  G.~G.~Ross,
  %``Models of fermion masses,''
%\href{http://www.slac.stanford.edu/spires/find/hep/www?irn=4868536}{SPIRES entry}
{\it Prepared for Theoretical Advanced Study Institute in Elementary Particle Physics (TASI 2000): Flavor Physics for the Millennium, Boulder,
Colorado, 4-30 Jun 2000}


%\cite{Froggatt:1978nt}
\bibitem{Froggatt:1978nt}
  C.~D.~Froggatt and H.~B.~Nielsen,
  %``Hierarchy Of Quark Masses, Cabibbo Angles And CP Violation,''
  Nucl.\ Phys.\  B {\bf 147} (1979) 277.
  %%CITATION = NUPHA,B147,277;%%

%\cite{Leurer:1992wg}
\bibitem{Leurer:1992wg}
  M.~Leurer, Y.~Nir and N.~Seiberg,
  %``Mass matrix models,''
  Nucl.\ Phys.\  B {\bf 398} (1993) 319
  [arXiv:hep-ph/9212278].
  %%CITATION = NUPHA,B398,319;%%

%\cite{Dine:1993np}
\bibitem{Dine:1993np}
  M.~Dine, R.~G.~Leigh and A.~Kagan,
  %``Flavor symmetries and the problem of squark degeneracy,''
  Phys.\ Rev.\  D {\bf 48} (1993) 4269
  [arXiv:hep-ph/9304299].
  %%CITATION = PHRVA,D48,4269;%%

%\cite{Kaplan:1993ej}
\bibitem{Kaplan:1993ej}
  D.~B.~Kaplan and M.~Schmaltz,
  %``Flavor unification and discrete nonAbelian symmetries,''
  Phys.\ Rev.\  D {\bf 49}, 3741 (1994)
  [arXiv:hep-ph/9311281].
  %%CITATION = PHRVA,D49,3741;%%

%\cite{Pomarol:1995xc}
\bibitem{Pomarol:1995xc}
  A.~Pomarol and D.~Tommasini,
  %``Horizontal symmetries for the supersymmetric flavor problem,''
  Nucl.\ Phys.\  B {\bf 466}, 3 (1996)
  [arXiv:hep-ph/9507462].
  %%CITATION = NUPHA,B466,3;%%

%\cite{Barbieri:1995uv}
\bibitem{Barbieri:1995uv}
  R.~Barbieri, G.~R.~Dvali and L.~J.~Hall,
  %``Predictions From A U(2) Flavour Symmetry In Supersymmetric Theories,''
  Phys.\ Lett.\  B {\bf 377}, 76 (1996)
  [arXiv:hep-ph/9512388].
  %%CITATION = PHLTA,B377,76;%%

%\cite{Binetruy:1996xk}
\bibitem{Binetruy:1996xk}
  P.~Binetruy, S.~Lavignac and P.~Ramond,
  %``Yukawa textures with an anomalous horizontal abelian symmetry,''
  Nucl.\ Phys.\  B {\bf 477} (1996) 353
  [arXiv:hep-ph/9601243].
  %%CITATION = NUPHA,B477,353;%%

%\cite{Dudas:1996fe}
\bibitem{Dudas:1996fe}
  E.~Dudas, C.~Grojean, S.~Pokorski and C.~A.~Savoy,
  %``Abelian flavour symmetries in supersymmetric models,''
  Nucl.\ Phys.\  B {\bf 481}, 85 (1996)
  [arXiv:hep-ph/9606383].
  %%CITATION = NUPHA,B481,85;%%

%\cite{Plentinger:2008nv}
\bibitem{Plentinger:2008nv}
  F.~Plentinger and G.~Seidl,
  %``Mapping out SU(5) GUTs with non-Abelian discrete flavor symmetries,''
  arXiv:0803.2889 [hep-ph].
  %%CITATION = ARXIV:0803.2889;%%

%\cite{Nir:1993mx}
\bibitem{Nir:1993mx}
  Y.~Nir and N.~Seiberg,
  %``Should squarks be degenerate?,''
  Phys.\ Lett.\  B {\bf 309}, 337 (1993)
  [arXiv:hep-ph/9304307].
  %%CITATION = PHLTA,B309,337;%%


%\cite{Leurer:1993gy}
\bibitem{Leurer:1993gy}
  M.~Leurer, Y.~Nir and N.~Seiberg,
  %``Mass matrix models: The Sequel,''
  Nucl.\ Phys.\  B {\bf 420} (1994) 468
  [arXiv:hep-ph/9310320].
  %%CITATION = NUPHA,B420,468;%%

%\cite{Nir:1996am}
\bibitem{Nir:1996am}
  Y.~Nir and R.~Rattazzi,
  %``Solving the Supersymmetric CP Problem with Abelian Horizontal Symmetries,''
  Phys.\ Lett.\  B {\bf 382} (1996) 363
  [arXiv:hep-ph/9603233].
  %%CITATION = PHLTA,B382,363;%%

\bibitem{Ross:2004qn}
G.~G.~Ross, L.~Velasco-Sevilla and O.~Vives,
%Spontaneous CP violation and non-Abelian family symmetry in SUSY
Nucl.\ Phys.\ B {\bf 692}, 50 (2004)
[arXiv:hep-ph/0401064]
 %%CITATION = NUPHA,B692,50;%%

%\cite{Joshipura:2000sn}
\bibitem{Joshipura:2000sn}
  A.~S.~Joshipura, R.~D.~Vaidya and S.~K.~Vempati,
  %``U(1) symmetry and R parity violation,''
  Phys.\ Rev.\  D {\bf 62}, 093020 (2000)
  [arXiv:hep-ph/0006138].
  %%CITATION = PHRVA,D62,093020;%%

\bibitem{WIP}
L.~Calibbi, J.~Jones~P\'erez, A. Masiero, J. Park and O.~Vives, work in progress.

%\cite{Dreiner:2003yr}
\bibitem{Dreiner:2003yr}
  H.~K.~Dreiner, H.~Murayama and M.~Thormeier,
  %``Anomalous flavor U(1)X for everything,''
  Nucl.\ Phys.\  B {\bf 729} (2005) 278
  [arXiv:hep-ph/0312012].
  %%CITATION = NUPHA,B729,278;%%

%\cite{Kane:2005va}
\bibitem{Kane:2005va}
  G.~L.~Kane, S.~F.~King, I.~N.~R.~Peddie and L.~Velasco-Sevilla,
  %``Study of theory and phenomenology of some classes of family symmetry  and
  %unification models,''
  JHEP {\bf 0508} (2005) 083
  [arXiv:hep-ph/0504038].
  %%CITATION = JHEPA,0508,083;%%

%\cite{Chankowski:2005qp}
\bibitem{Chankowski:2005qp}
  P.~H.~Chankowski, K.~Kowalska, S.~Lavignac and S.~Pokorski,
  %``Update on fermion mass models with an anomalous horizontal U(1)
  %symmetry,''
  Phys.\ Rev.\  D {\bf 71} (2005) 055004
  [arXiv:hep-ph/0501071].
  %%CITATION = PHRVA,D71,055004;%%

%\cite{Nir:2002ah}
\bibitem{Nir:2002ah}
  Y.~Nir and G.~Raz,
  %``Quark squark alignment revisited,''
  Phys.\ Rev.\  D {\bf 66} (2002) 035007
  [arXiv:hep-ph/0206064].
  %%CITATION = PHRVA,D66,035007;%%


%\cite{Barbieri:1996ww}
\bibitem{Barbieri:1996ww}
  R.~Barbieri, L.~J.~Hall, S.~Raby and A.~Romanino,
  %``Unified theories with U(2) flavor symmetry,''
  Nucl.\ Phys.\  B {\bf 493}, 3 (1997)
  [arXiv:hep-ph/9610449].
  %%CITATION = NUPHA,B493,3;%%


\bibitem{King:2001uz}
S.~F.~King and G.~G.~Ross,
%Fermion masses and mixing angles from SU(3) family symmetry
Phys.\ Lett.\ B {\bf 520}, 243 (2001)
[arxiv:hep-ph/0108112]
%%CITATION = PHLTA,B520,243;%%


\bibitem{King:2003rf}
S.~F.~King and G.~G.~Ross,
%Fermion masses and mixing angles from SU(3) family symmetry and  unification
Phys.\ Lett.\ B {\bf 574}, 239 (2003)
[arxiv:hep-ph/0307190]
%%CITATION = PHLTA,B574,239;%%

%\cite{deMedeirosVarzielas:2005ax}
\bibitem{deMedeirosVarzielas:2005ax}
  I.~de Medeiros Varzielas and G.~G.~Ross,
  %``SU(3) family symmetry and neutrino bi-tri-maximal mixing,''
  Nucl.\ Phys.\  B {\bf 733} (2006) 31
  [arXiv:hep-ph/0507176].
  %%CITATION = NUPHA,B733,31;%%

%\cite{Babu:2002dz}
\bibitem{Babu:2002dz}
  K.~S.~Babu, E.~Ma and J.~W.~F.~Valle,
  %``Underlying A(4) symmetry for the neutrino mass matrix and the quark  mixing
  %matrix,''
  Phys.\ Lett.\  B {\bf 552} (2003) 207
  [arXiv:hep-ph/0206292].
  %%CITATION = PHLTA,B552,207;%%

%\cite{Hirsch:2003dr}
\bibitem{Hirsch:2003dr}
  M.~Hirsch, J.~C.~Romao, S.~Skadhauge, J.~W.~F.~Valle and A.~Villanova del Moral,
  %``Phenomenological tests of supersymmetric A(4) family symmetry model of
  %neutrino mass,''
  Phys.\ Rev.\  D {\bf 69} (2004) 093006
  [arXiv:hep-ph/0312265].
  %%CITATION = PHRVA,D69,093006;%%


%\cite{Altarelli:2005yx}
\bibitem{Altarelli:2005yx}
  G.~Altarelli and F.~Feruglio,
  %``Tri-bimaximal neutrino mixing, A(4) and the modular symmetry,''
  Nucl.\ Phys.\  B {\bf 741} (2006) 215
  [arXiv:hep-ph/0512103].
  %%CITATION = NUPHA,B741,215;%%

%\cite{deMedeirosVarzielas:2005qg}
\bibitem{deMedeirosVarzielas:2005qg}
  I.~de Medeiros Varzielas, S.~F.~King and G.~G.~Ross,
  %``Tri-bimaximal neutrino mixing from discrete subgroups of SU(3) and  SO(3)
  %family symmetry,''
  Phys.\ Lett.\  B {\bf 644} (2007) 153
  [arXiv:hep-ph/0512313].
  %%CITATION = PHLTA,B644,153;%%
%\cite{Ma:2006sk}
\bibitem{Ma:2006sk}
  E.~Ma, H.~Sawanaka and M.~Tanimoto,
  %``Quark masses and mixing with A(4) family symmetry,''
  Phys.\ Lett.\  B {\bf 641} (2006) 301
  [arXiv:hep-ph/0606103].
  %%CITATION = PHLTA,B641,301;%%
 
%\cite{Ma:2006ip}
\bibitem{Ma:2006ip}
  E.~Ma,
  %``Neutrino mass matrix from Delta(27) symmetry,''
  Mod.\ Phys.\ Lett.\  A {\bf 21} (2006) 1917
  [arXiv:hep-ph/0607056].
  %%CITATION = MPLAE,A21,1917;%%

%\cite{deMedeirosVarzielas:2006fc}
\bibitem{deMedeirosVarzielas:2006fc}
  I.~de Medeiros Varzielas, S.~F.~King and G.~G.~Ross,
  %``Neutrino tri-bi-maximal mixing from a non-Abelian discrete family
  %symmetry,''
  Phys.\ Lett.\  B {\bf 648} (2007) 201
  [arXiv:hep-ph/0607045].
  %%CITATION = PHLTA,B648,201;%%

%\cite{Feruglio:2007uu}
\bibitem{Feruglio:2007uu}
  F.~Feruglio, C.~Hagedorn, Y.~Lin and L.~Merlo,
  %``Tri-bimaximal neutrino mixing and quark masses from a discrete flavour
  %symmetry,''
  Nucl.\ Phys.\  B {\bf 775} (2007) 120
  [arXiv:hep-ph/0702194].
  %%CITATION = NUPHA,B775,120;%%


\bibitem{Roberts:2001}
R.~G.~Roberts, A.~Romanino, G.~G.~Ross and L.~Velasco-Sevilla,
%``Precision test of a fermion mass texture,''
Nucl.\ Phys.\ B {\bf 615} (2001) 358
[arXiv:hep-ph/0104088].
%%CITATION = HEP-PH 0104088;%%

%\cite{Georgi:1979df}
\bibitem{Georgi:1979df}
  H.~Georgi and C.~Jarlskog,
  %``A New Lepton - Quark Mass Relation In A Unified Theory,''
  Phys.\ Lett.\  B {\bf 86} (1979) 297.
  %%CITATION = PHLTA,B86,297;%%

\bibitem{Ross:2002mr}
G.~G.~Ross and O.~Vives,
%Yukawa structure, flavour and CP violation in supergravity
Phys.\ Rev.\ D {\bf 67}, 095013 (2003)
[arXiv:hep-ph/0211279]

%\cite{Antusch:2007re}
\bibitem{Antusch:2007re}
  S.~Antusch, S.~F.~King and M.~Malinsky,
  %``Solving the SUSY Flavour and CP Problems with SU(3) Family Symmetry,''
  arXiv:0708.1282 [hep-ph].
  %%CITATION = ARXIV:0708.1282;%%

%\cite{Olive:2008vv}
\bibitem{Olive:2008vv}
  K.~A.~Olive and L.~Velasco-Sevilla,
  %``Constraints on Supersymmetric Flavour Models from b->s gamma,''
  arXiv:0801.0428 [hep-ph].
  %%CITATION = ARXIV:0801.0428;%%

%\cite{Brignole:1993dj}
\bibitem{Brignole:1993dj}
  A.~Brignole, L.~E.~Ibanez and C.~Munoz,
  %``Towards a theory of soft terms for the supersymmetric Standard Model,''
  Nucl.\ Phys.\  B {\bf 422}, 125 (1994)
  [Erratum-ibid.\  B {\bf 436}, 747 (1995)]
  [arXiv:hep-ph/9308271].
  %%CITATION = NUPHA,B422,125;%%

\bibitem{Giudice:1998bp}
       G.~F.~Giudice and R.~Rattazzi,
        %``Theories with gauge-mediated supersymmetry breaking,''
        Phys.\ Rept.\  {\bf 322} (1999) 419
        [arXiv:hep-ph/9801271].
        %%CITATION = PRPLC,322,419;%%

%\cite{Randall:1998uk}
\bibitem{Randall:1998uk}
  L.~Randall and R.~Sundrum,
  %``Out of this world supersymmetry breaking,''
  Nucl.\ Phys.\  B {\bf 557}, 79 (1999)
  [arXiv:hep-th/9810155].
  %%CITATION = NUPHA,B557,79;%%

%\cite{Giudice:1998xp}
\bibitem{Giudice:1998xp}
  G.~F.~Giudice, M.~A.~Luty, H.~Murayama and R.~Rattazzi,
  %``Gaugino mass without singlets,''
  JHEP {\bf 9812}, 027 (1998)
  [arXiv:hep-ph/9810442].
  %%CITATION = JHEPA,9812,027;%%

%\cite{Pomarol:1999ie}
\bibitem{Pomarol:1999ie}
  A.~Pomarol and R.~Rattazzi,
  %``Sparticle masses from the superconformal anomaly,''
  JHEP {\bf 9905}, 013 (1999)
  [arXiv:hep-ph/9903448].
  %%CITATION = JHEPA,9905,013;%%

%\cite{King:2004tx}
\bibitem{King:2004tx}
  S.~F.~King, I.~N.~R.~Peddie, G.~G.~Ross, L.~Velasco-Sevilla and O.~Vives,
  %``Kaehler corrections and softly broken family symmetries,''
  JHEP {\bf 0507} (2005) 049
  [arXiv:hep-ph/0407012].
  %%CITATION = JHEPA,0507,049;%%


\bibitem{CurreEDM}
B.~C.~Regan and E.~D.~Commins and C.~J.~Schmidt and D.~DeMille,
% New limit on the electron electric dipole moment
Phys.\ Rev.\ Lett.\ {\bf 88} (2002) 071805
%%CITATION = PRLTA,88,071805;%%

\bibitem{CurrmEDM}
J.~Bailey {\it et al.},
% New Limits on the Electric Dipole Moment of Positive and Negative Muons
J.\ Phys.\ G {\bf 4} (1978) 345
%%CITATION = JPHGB,G4,345;%%


\bibitem{CurrtEDM}
K.~Inami {\it et al.},
% Search for the electric dipole moment of the tau lepton
Phys.\ Lett.\ B {\bf 551} (2003) 16
[arXiv:hep-ex/0210066]
%%CITATION = PHLTA,B551,16;%%


\bibitem{CurrnEDM}
C.~A.~Baker {\it et al.},
% An improved experimental limit on the electric dipole moment of the neutron
Phys.\ Rev.\ Lett.\ {\bf 97} (2006) 131801
%%CITATION = PRLTA,97,131801;%%

\bibitem{FuteEDM}
 S.~K.~Lamoreaux,
% Solid state systems for electron electric dipole moment and other fundamental measurements
[arXiv:nucl-ex/0109014]
 %%CITATION = NUCL-EX/0109014;%%

\bibitem{FutmEDM}
 A.~Adelmann and K.~Kirch,
% Search for the muon electric dipole moment using a compact storage  ring
[arXiv:hep-ex/0606034]
%%CITATION = HEP-EX/0606034;%%

\bibitem{FuttEDM}
 G.~A.~Gonzalez-Sprinberg, J.~Bernabeu and J.~Vidal,
  %``$\tau$ electric dipole moment with polarized beams,''
  arXiv:0707.1658 [hep-ph].
  %%CITATION = ARXIV:0707.1658;%%


\bibitem{FutnEDM}
http://p25ext.lanl.gov/edm/edm.html

\bibitem{SMeEDM}
M.~E.~Pospelov and  I.~B.~Khriplovich,
% Electric dipole moment of the W boson and the electron in the Kobayashi-Maskawa model
Sov.\ J.\ Nucl.\ Phys.\ {\bf 53} (1991) 638


%\cite{Ellis:1982tk}
\bibitem{Ellis:1982tk}
  J.~R.~Ellis, S.~Ferrara and D.~V.~Nanopoulos,
  %``CP Violation And Supersymmetry,''
  Phys.\ Lett.\  B {\bf 114}, 231 (1982).
  %%CITATION = PHLTA,B114,231;%%

%\cite{Buchmuller:1982ye}
\bibitem{Buchmuller:1982ye}
  W.~Buchmuller and D.~Wyler,
   ``CP Violation And R Invariance In Supersymmetric Models Of Strong And
  %Electroweak Interactions,''
  Phys.\ Lett.\  B {\bf 121}, 321 (1983).
  %%CITATION = PHLTA,B121,321;%%


%\cite{Polchinski:1983zd}
\bibitem{Polchinski:1983zd}
J.~Polchinski and M.~B.~Wise,
%``The Electric Dipole Moment Of The Neutron In Low-Energy Supergravity,''
Phys.\ Lett.\  B {\bf 125}, 393 (1983).
%%CITATION = PHLTA,B125,393;%%


%\cite{Franco:1983xm}
\bibitem{Franco:1983xm}
  E.~Franco and M.~L.~Mangano,
  %``CP And Flavor Nonconservation In Softly Broken Supersymmetric Theories,''
  Phys.\ Lett.\  B {\bf 135}, 445 (1984).
  %%CITATION = PHLTA,B135,445;%%

%\cite{Dugan:1984qf}
\bibitem{Dugan:1984qf}
  M.~Dugan, B.~Grinstein and L.~J.~Hall,
  %``CP Violation In The Minimal N=1 Supergravity Theory,''
  Nucl.\ Phys.\  B {\bf 255}, 413 (1985).
  %%CITATION = NUPHA,B255,413;%%



%\cite{Nath:1991dn}
\bibitem{Nath:1991dn}
P.~Nath,
%``CP Violation via electroweak gauginos and the electric dipole moment of the
%electron,''
Phys.\ Rev.\ Lett.\  {\bf 66}, 2565 (1991).
%%CITATION = PRLTA,66,2565;%%

%\cite{Fischler:1992ha}
\bibitem{Fischler:1992ha}
  W.~Fischler, S.~Paban and S.~D.~Thomas,
   ``Bounds on microscopic physics from P and T violation in atoms and
  %molecules,''
  Phys.\ Lett.\  B {\bf 289}, 373 (1992)
  [arXiv:hep-ph/9205233].
  %%CITATION = PHLTA,B289,373;%%

\bibitem{Ayazi:2007kd}
  S.~Y.~Ayazi and Y.~Farzan,
  %``Electron electric dipole moment from lepton flavor violation,''
  JHEP {\bf 0706}, 013 (2007)
  [arXiv:hep-ph/0702149].
  %%CITATION = JHEPA,0706,013;%%

%\bibitem{SMnEDM}
%I.~B.~Khriplovich and A.~R.~Zhitnitsky,
%% What Is The Value Of The Neutron Electric Dipole Moment In The Kobayashi-Maskawa Model?
%Phys.\ Lett.\ B {\bf 109} (1982) 490

%\cite{Ibrahim:1998je}
\bibitem{Ibrahim:1998je}
T.~Ibrahim and P.~Nath,
%``The neutron and the lepton EDMs in MSSM, large CP violating phases, and
%the cancellation mechanism,''
Phys.\ Rev.\  D {\bf 58}, 111301 (1998)
[Erratum-ibid.\  D {\bf 60}, 099902 (1999)]
[arXiv:hep-ph/9807501].
%%CITATION = PHRVA,D58,111301;%%

\bibitem{Giudice:1988yz}
G.~F.~Giudice and A.~Masiero
%A Natural Solution to the mu Problem in Supergravity Theories
Phys.\ Lett.\ B {\bf 206}, 480 (1988)

\bibitem{Barr:1988wk}
S.~Barr and A.~Masiero,
%SPONTANEOUS CP VIOLATION IN THEORIES WITH LOW-ENERGY SUPERSYMMETRY
Phys.\ Rev.\ D {\bf 38}, 366 (1988)

%\cite{Gabbiani:1996hi}
\bibitem{Gabbiani:1996hi}
F.~Gabbiani, E.~Gabrielli, A.~Masiero and L.~Silvestrini,
%``A complete analysis of FCNC and CP constraints in general SUSY extensions
%of the standard model,''
Nucl.\ Phys.\  B {\bf 477}, 321 (1996)
[arXiv:hep-ph/9604387].
%%CITATION = NUPHA,B477,321;%%

%\cite{Hagelin:1992tc}
\bibitem{Hagelin:1992tc}
J.~S.~Hagelin, S.~Kelley and T.~Tanaka,
%``Supersymmetric flavor changing neutral currents: Exact amplitudes and
%phenomenological analysis,''
Nucl.\ Phys.\  B {\bf 415} (1994) 293.
%%CITATION = NUPHA,B415,293;%%

\bibitem{Masina:2002mv}
I.~Masina and C.~A.~Savoy,
%Sleptonarium (constraints on the CP and flavour pattern of scalar lepton masses)
Nucl.\ Phys.\ B {\bf 661} (2003) 365
[arxiv:hep-ph/0211283]

\bibitem{Borzumati:1986qx}
F.~Borzumati and A.~Masiero,
%Large Muon and electron Number Violations in Supergravity Theories
Phys.\ Rev.\ Lett.\ {\bf 57} (1986) 961

%\cite{Casas:2001sr}
\bibitem{Casas:2001sr}
J.~A.~Casas and A.~Ibarra,
%``Oscillating neutrinos and mu --> e, gamma,''
Nucl.\ Phys.\  B {\bf 618}, 171 (2001)
[arXiv:hep-ph/0103065].
%%CITATION = NUPHA,B618,171;%%

%\cite{Masiero:2002jn}
\bibitem{Masiero:2002jn}
A.~Masiero, S.~K.~Vempati and O.~Vives,
%``Seesaw and lepton flavour violation in SUSY SO(10),''
Nucl.\ Phys.\  B {\bf 649}, 189 (2003)
[arXiv:hep-ph/0209303].
%%CITATION = NUPHA,B649,189;%%

%\cite{Abel:2001vy}
\bibitem{Abel:2001vy}
S.~Abel, S.~Khalil and O.~Lebedev,
%``EDM constraints in supersymmetric theories,''
Nucl.\ Phys.\  B {\bf 606}, 151 (2001)
[arXiv:hep-ph/0103320].
%%CITATION = NUPHA,B606,151;%%

%\cite{Bartl:2003ju}
\bibitem{Bartl:2003ju}
  A.~Bartl, W.~Majerotto, W.~Porod and D.~Wyler,
  %``Effect of supersymmetric phases on lepton dipole moments and rare  lepton
  %decays,''
  Phys.\ Rev.\  D {\bf 68}, 053005 (2003)
  [arXiv:hep-ph/0306050].
  %%CITATION = PHRVA,D68,053005;%%

%\cite{Botella:2004ks}
\bibitem{Botella:2004ks}
F.~J.~Botella, M.~Nebot and O.~Vives,
%``Invariant approach to flavour-dependent CP-violating phases in the  MSSM,''
JHEP {\bf 0601}, 106 (2006)
[arXiv:hep-ph/0407349].
%%CITATION = JHEPA,0601,106;%%

\bibitem{Bartl:1999bc}
A.~Bartl, T.~Gajdosik, W.~Porod, P.~Stockinger and H.~Stremnitzer,
%Electron and neutron electric dipole moments in the constrained MSSM
Phys.\ Rev.\ D {\bf 60}, 073003 (1999)
[arXiv:hep-ph/9903402]

%\cite{Bartl:2001wc}
\bibitem{Bartl:2001wc}
  A.~Bartl, T.~Gajdosik, E.~Lunghi, A.~Masiero, W.~Porod, H.~Stremnitzer and O.~Vives,
  %``General flavor blind MSSM and CP violation,''
  Phys.\ Rev.\  D {\bf 64} (2001) 076009
  [arXiv:hep-ph/0103324].
  %%CITATION = PHRVA,D64,076009;%%

%\cite{Hisano:1995cp}
\bibitem{Hisano:1995cp}
J.~Hisano, T.~Moroi, K.~Tobe and M.~Yamaguchi,
%``Lepton-Flavor Violation via Right-Handed Neutrino Yukawa Couplings in
%Supersymmetric Standard Model,''
Phys.\ Rev.\  D {\bf 53}, 2442 (1996)
[arXiv:hep-ph/9510309].
%%CITATION = PHRVA,D53,2442;%%

%\cite{Paradisi:2005fk}
\bibitem{Paradisi:2005fk}
  P.~Paradisi,
  %``Constraints on SUSY lepton flavour violation by rare processes,''
  JHEP {\bf 0510}, 006 (2005)
  [arXiv:hep-ph/0505046].
  %%CITATION = JHEPA,0510,006;%%



\bibitem{Ciuchini:2007ha}
M.~Ciuchini {\it et al.},
%Soft SUSY breaking grand unification: Leptons vs quarks on the flavor playground
Nucl.\ Phys.\ B {\bf 783} (2007) 112
[arxiv:hep-ph/0702144]

\bibitem{Aubert:2005wa}
  B.~Aubert {\it et al.}  [BABAR Collaboration],
  %``Search for lepton flavor violation in the decay $\tau^\pm \to e^\pm
  %\gamma$,''
  Phys.\ Rev.\ Lett.\  {\bf 96} (2006) 041801
  [arXiv:hep-ex/0508012].
  %%CITATION = PRLTA,96,041801;%%

\bibitem{Ahmed:2001eh}
  M.~Ahmed {\it et al.}  [MEGA Collaboration],
  %``Search for the lepton-family-number nonconserving decay mu+ --> e+
  %gamma,''
  Phys.\ Rev.\  D {\bf 65} (2002) 112002
  [arXiv:hep-ex/0111030].
  %%CITATION = PHRVA,D65,112002;%%

\bibitem{meg}
Web page: http://meg.psi.ch.

\bibitem{Banerjee:2007rj}
S.~Banerjee,
%Searches for lepton flavor violating decays tau+- -/-> l+- gamma, tau+- --> l+- P0 (where l- = e-, mu-, and P0 = pi0, eta, eta') at B-factories: Status and combinations
Nucl.\ Phys.\ Proc.\ Suppl.\ {\bf 169} (2007) 199
[arxiv:hep-ex/0702017]

\bibitem{Lusiani:2007cb}
A.~Lusiani,
%Experimental Review on Lepton Universality and Lepton Flavour Violation tests at the B-factories
[arXiv:0709.1599 [hep-ex]]

%\cite{Petcov:2003zb}
\bibitem{Petcov:2003zb}
S.~T.~Petcov, S.~Profumo, Y.~Takanishi and C.~E.~Yaguna,
%``Charged lepton flavor violating decays: Leading logarithmic  approximation
%versus full RG results,''
Nucl.\ Phys.\  B {\bf 676}, 453 (2004)
[arXiv:hep-ph/0306195].
%%CITATION = NUPHA,B676,453;%%

%\cite{Yao:2006px}
\bibitem{Yao:2006px}
W.~M.~Yao {\it et al.}  [Particle Data Group],
%``Review of particle physics,''
J.\ Phys.\ G {\bf 33} (2006) 1.
%%CITATION = JPHGB,G33,1;%%

\bibitem{Bona:2007qt}
  M.~Bona {\it et al.},
  %``SuperB: A High-Luminosity Asymmetric e+ e- Super Flavor Factory. Conceptual
  %Design Report,''
  arXiv:0709.0451 [hep-ex].
  %%CITATION = ARXIV:0709.0451;%%

\end{thebibliography}
\end{document}